\documentclass[pre, twocolumn, floatfix, groupedaddress]{revtex4}%
\usepackage{amsfonts}
\usepackage{amsmath}
\usepackage{amssymb}
\usepackage{graphicx}
\usepackage{bm}
\usepackage{subfigure}
\usepackage{color}
\usepackage{booktabs}
\usepackage[colorlinks=true,linkcolor=blue,linktocpage=true,
urlcolor=blue,citecolor=blue,anchorcolor=blue]{hyperref}%
\providecommand{\U}[1]{\protect\rule{.1in}{.1in}}
\begin{document}
\title{Modulation of chiral anomaly and bilinear magnetoconductivity in Weyl semimetals by impurity-resonance states}
\author{Mei-Wei Hu$^{1}$}
\author{Zhuo-Yan Fang$^{1}$}
\author{Hou-Jian Duan$^{1,2}$}
\author{Mou Yang$^{1,2}$}
\author{Ming-Xun Deng$^{1,2}$}
\email{dengmingxun@scnu.edu.cn}
\author{Rui-Qiang Wang$^{1,2}$}
\email{wangruiqiang@m.scnu.edu.cn}
\affiliation{$^{1}$Guangdong Basic Research Center of Excellence for Structure and
Fundamental Interactions of Matter, Guangdong Provincial Key Laboratory of
Quantum Engineering and Quantum Materials, School of Physics, South China
Normal University, Guangzhou 510006, China}
\affiliation{$^{2}$Guangdong-Hong Kong Joint Laboratory of Quantum Matter, Frontier
Research Institute for Physics, South China Normal University, Guangzhou
510006, China}

\begin{abstract}
The phenomenon of nonlinear transport has attracted tremendous interest within
the condensed matter community. We present a theoretical framework for nonlinear transport based on the nonequilibrium retarded Green's function, and examine the impact of disorder on nonlinear magnetotransport in Weyl semimetals (WSMs). It is demonstrated that bilinear magnetoconductivity can be induced in disordered WSMs by several mechanisms, including impurity-induced tilting of the Weyl cones, Lorentz-force-induced normal orbital magnetic moment, and chiral anomaly arising from the Berry-curvature-induced anomalous orbital magnetic moment. Additionally, we observe that the localization of Weyl fermions by impurity scattering will lead to resonant dips in both the chiral chemical potential and magnetoconductivity when the Fermi energy approaches the impurity resonance states. Our findings offer a theoretical proposition for modulating nonreciprocal transport in topological semimetals.
\end{abstract}
\maketitle

\section{Introduction}

Weyl semimetals (WSMs) are members of the topological semimetal family,
well-known for hosting low-energy Weyl fermion excitation with well-defined chirality near the Weyl
nodes\cite{PhysRevB.83.205101,PhysRevLett.107.127205,PhysRevX.5.011029,Pesin:2012aa,Katsnelson:2006aa,Yan:2017aa,RevModPhys.90.015001,RevModPhys.88.035005,RevModPhys.93.025002}%
. The Weyl nodes always come in pairs of opposite chiralities connected by
nonclosed Fermi arc surface states in momentum space, acting as sources and sinks of the Berry curvature. WSMs can be classified into two categories: inversion symmetry-broken WSMs and time-reversal symmetry-broken WSMs, with the Weyl nodes separated in momentum and energy space,
respectively. One of the most distinctive features of WSMs is the chiral
anomaly, i.e., a population difference between opposite-chirality Weyl valleys can be induced by nonorthogonal electric and magnetic
fields\cite{PhysRevLett.113.247203,semi_PhysRevB.88.104412,semi_PhysRevB.95.245128,semi_PhysRevLett.119.176804}%
. The chiral anomaly can give rise to various observable phenomena, such as
the negative magnetoresistance\cite{Xiong413,Li:2015aa,PhysRevLett.122.036601,PhysRevB.99.165146}%
, quantum oscillations\cite{Nat.Commun.5.5161,Moll2016} and topological pumping
effects\cite{PhysRevB.106.075139}.

In recent studies, there have been proposals for chiral-anomaly-induced nonlinear transport phenomena  in WSMs, such as the nonlinear Hall effect\cite{im_PhysRevB.103.045105,semi_PhysRevB.106.035423,im_PhysRevB.103.245119}. The nonlinear Hall effect, initially introduced by Sodemann and Fu\cite{Fu_PhysRevLett.115.216806}, is a new member of the Hall family and has attracted great interest both theoretically and experimentally in recent
years\cite{Juan:2017aa,Xu:2018aa,Kang:2019aa,PhysRevLett.121.266601,PhysRevLett.127.277201,PhysRevLett.127.277202,PhysRevLett.129.086602,PhysRevLett.130.126303,PhysRevB.108.L241104}. Unlike its linear counterpart\cite{RevModPhys.82.1539,RevModPhys.82.1959}, the nonlinear Hall effect typically requires the breaking of inversion symmetry instead of time-reversal symmetry. The nonlinear Hall effect has been observed in inversion symmetry-broken WSMs\cite{Ma:2019aa}. Additionally, the nonlinear anomalous Hall effect induced by the Berry curvature dipole has also been realized in
WSMs\cite{Fu_PhysRevLett.115.216806,semi_PhysRevB.94.245121,PhysRevLett.123.246602}. In addition to the intrinsic contributions, extrinsic mechanisms, such as the
nonlinear side jump and
skew-scattering\cite{PhysRevB.99.155404,im_PhysRevB.100.195117,Du:2019aa}, can
also contribute to the nonlinear transport. Furthermore, besides giving rise to
nontrivial nonlinear responses in the weak magnetic field regime, the
nonlinear conductivities can exhibit quantum oscillation behavior owing to the
Landau level quantization in the strong magnetic field
limit\cite{QM_PhysRevB.107.L081107}. More interestingly, when the nonlinear
effects become significant for the applied external field getting larger, the violation of reciprocity principles elucidated by Onsager\cite{PhysRev.38.2265}
can take place, resulting in nonreciprocal transport behavior underpin
the functionality of key electronic devices like diodes and
photodetectors\cite{Fruchart:2021aa,doi:10.1126/science.aaz9146}.

 The previous theoretical studies were mostly based on the Boltzmann theory,
where the electric and magnetic fields are treated perturbatively and incorporated into transport via the semiclassical equations of
motion\cite{WP_PhysRevB.59.14915,RevModPhys.82.1959}. This framework, extensively utilized for exploring transport properties in topological materials\cite{PhysRevLett.113.247203,semi_PhysRevB.88.104412,semi_PhysRevB.95.245128,semi_PhysRevLett.119.176804,semi_PhysRevB.94.245121,PhysRevLett.123.246602,PhysRevB.99.155404,im_PhysRevB.100.195117,Du:2019aa}, provides an intuitive understanding of topological transport phenomena. However, the Berry curvature within the semiclassical equations becomes ill-defined at the gap-closing points. Consequently, both linear and nonlinear magnetoconductivities predicted by the semiclassical theory exhibit unphysical divergence as the Fermi energy approaches the Weyl nodes\cite{im_PhysRevB.103.045105,semi_PhysRevB.106.035423,im_PhysRevB.103.245119}. This contrasts with the results obtained from the quantum Boltzmann theory in strong magnetic field regimes\cite{PhysRevB.99.165146,PhysRevLett.122.036601,QM_PhysRevB.107.L081107}. 
A unified theory across the weak and strong magnetic field limits is still lacking for nonlinear transport. Besides, the Boltzmann equation relies on phenomenological relaxation processes to determine the distribution function, where the constant relaxation time approximation is often employed\cite{PhysRevLett.113.247203,semi_PhysRevB.88.104412,semi_PhysRevB.95.245128,semi_PhysRevLett.119.176804,semi_PhysRevB.94.245121,PhysRevLett.123.246602,PhysRevB.99.155404,im_PhysRevB.100.195117,Du:2019aa}.  In real disordered systems, impurity scattering processes can lead not only to quasimomentum and energy relaxations but also to the emergence of impurity resonance states\cite{PhysRevB.94.235116,PhysRevB.96.155141,PhysRevB.99.165146}. The constant relaxation time approximation falls short in describing these states.

In this work, we develop a unified theory based on the retarded Green's
function and explore the nonlinear magnetotransport in disordered
WSMs,  where the impurity-resonance states are taken into account. We find that
bilinear magnetoconductivity that is  linearly dependent on both electric and magnetic fields can be induced by several mechanisms in disordered WSMs,
 including impurity-induced tilting of the Weyl cones, Lorentz-force induced orbital magnetic moment, and the Berry-curvature induced anomalous orbital magnetic moment. We identify three key features in the second-order nonlinear conductivity: ($\mathrm{i}$) The nonlinear anomalous Hall conductivity and orbital-magnetic-moment induced nonlinear Hall conductivity exhibit a linear dependence on lifetime, while other nonlinear conductivities, such as the Drude conductivity, show a quadratic dependence.
($\mathrm{ii}$) A finite quadratic-in-lifetime nonlinear conductivity requires at least two identical subscripts, whereas linear-in-lifetime nonlinear conductivities require at least two different subscripts. ($\mathrm{iii}$) The chiral anomaly can induce both the linear- and quadratic-in-lifetime nonlinear magnetoconductivity, with the former
(latter) being a constant (vanishing) for $E_{F}\rightarrow0$ and becoming
$B/E_{F}^{2}$ ($B/E_{F}$) dependent for $|E_{F}|\gg1$. This contrasts with the
$B^{2}/E_{F}^{2}$ dependence in the linear case, where $E_{F}$ is the Fermi
energy. Additionally, due to the impurity-induced resonance states, resonant dips are observed in both chiral anomaly and nonlinear conductivities with variations in the Fermi energy. 

The rest of this paper is organized as follows. In Sec. \ref{TBG}, we present the theory based on the retarded Green's function and derive the corresponding formula for current density in Sec. \ref{CDF}. Subsequently, in Sec. \ref{NRG}, we derive the nonequilibrium retarded Green's function for disordered WSMs, followed by a discussion on the current-induced anomalous orbital magnetic moment in Sec. \ref{CAM}. The chiral anomaly and nonlinear magnetoconductivity in disordered WSMs are analyzed in Secs. \ref{CFM} and \ref{CBM}, respectively. Finally, a concise summary is provided in the last section.

\section{Theory based on the retarded Green's function}

\label{TBG} We start from a general system subjected to external fields, which
can be described by the Hamiltonian
\begin{equation}
H=\int d^{3}\boldsymbol{r}\Psi^{\dag}(\boldsymbol{r})\mathcal{H}%
(\boldsymbol{r})\Psi(\boldsymbol{r}), \label{eq_Hr}%
\end{equation}
where $\mathcal{H}(\boldsymbol{r})=\mathcal{\hat{H}}_{0}+\hat{V}%
(\boldsymbol{r})+\mathcal{\hat{H}}_{\mathrm{ex}}\left(  \boldsymbol{r}\right)
$ is the local single-particle Hamiltonian and $\Psi(\boldsymbol{r})$ is the
fermion annihilation operator at position $\boldsymbol{r}$. Specifically,
$\mathcal{\hat{H}}_{0}$ possesses the spatial translation symmetry which
governs the energy-band quasiparticles. $\hat{V}(\boldsymbol{r})=\sum
_{n}V_{n}\delta\left(  \boldsymbol{r}-\boldsymbol{R}_{n}\right)  $ denotes the
scattering potential by impurities or defects. $\mathcal{\hat{H}%
}_{\mathrm{ex}}\left(  \boldsymbol{r}\right)  =q\Phi-\hat{\boldsymbol{J}}%
\cdot\boldsymbol{A}-\hat{\boldsymbol{M}}_{s}\cdot\boldsymbol{B}$ couples the
local charge $q$, magnetic moment $\hat{\boldsymbol{M}}_{s}$ and current
density $\hat{\boldsymbol{J}}=q\dot{\boldsymbol{r}}$ to external fields. Here, the
external electric and magnetic fields can be denoted by $\boldsymbol{E}%
=-\boldsymbol{\nabla}\Phi-\partial_{t}\boldsymbol{A}$ and $\boldsymbol{B}%
=\boldsymbol{\nabla}\times\boldsymbol{A}$, in which $\Phi$ and $\boldsymbol{A}%
$ are the electromagnetic scalar and vector potentials, respectively.

Provided the external fields are homogeneous in real space, we can rewrite
$\mathcal{\hat{H}}_{\mathrm{ex}}\left(  \boldsymbol{r}\right)
=-\boldsymbol{r}\cdot\hat{\boldsymbol{F}}-\hat{\boldsymbol{M}}_{s}%
\cdot\boldsymbol{B}$ with the generalized force defined
as\cite{Goldstein2002,PhysRevB.106.075139}
\begin{equation}
\hat{\boldsymbol{F}}=-\frac{\partial\mathcal{\hat{H}}_{\mathrm{ex}}\left(
\boldsymbol{r}\right)  }{\partial\boldsymbol{r}}=-\boldsymbol{\nabla
}\mathcal{\hat{H}}_{\mathrm{ex}}\left(  \boldsymbol{r}\right)  +\frac{d}%
{dt}\frac{\partial\mathcal{\hat{H}}_{\mathrm{ex}}\left(  \boldsymbol{r}%
\right)  }{\partial\dot{\boldsymbol{r}}}. \label{eq_pp}%
\end{equation}
Although the translational symmetry is broken by the external fields,
the plane waves, as eigenstates of the momentum operator, can form a complete
system. Thus, we can expand $\Psi(\boldsymbol{r})=\sum_{\boldsymbol{k}%
}c_{\boldsymbol{k}}e^{i\boldsymbol{k}\cdot\boldsymbol{r}}$ and project Eq.
(\ref{eq_Hr}) onto momentum space, yielding%
\begin{equation}
H=\sum_{\boldsymbol{k}}c_{\boldsymbol{k}}^{\dag}\mathcal{H}_{\boldsymbol{k}%
}c_{\boldsymbol{k}}+\sum_{\boldsymbol{kk}^{\prime}}c_{\boldsymbol{k}}^{\dag
}\left(  \mathcal{V}_{\boldsymbol{kk}^{\prime}}+\mathcal{W}_{\boldsymbol{kk}%
^{\prime}}\right)  c_{\boldsymbol{k}^{\prime}}, \label{eq_Hkk}%
\end{equation}
where $\mathcal{H}_{\boldsymbol{k}}=e^{-i\boldsymbol{k}\cdot\boldsymbol{r}%
}\mathcal{\hat{H}}_{0}e^{i\boldsymbol{k}\cdot\boldsymbol{r}}$ and
\begin{align}
\mathcal{V}_{\boldsymbol{kk}^{\prime}}  &  =\sum_{n}V_{n}e^{-i\left(
\boldsymbol{k}-\boldsymbol{k}^{\prime}\right)  \cdot\boldsymbol{R}_{n}},\\
\mathcal{W}_{\boldsymbol{kk}^{\prime}}  &  =\hat{U}_{\boldsymbol{kk}^{\prime}%
}-\hat{\boldsymbol{M}}_{s}\cdot\boldsymbol{B}\delta_{\boldsymbol{kk}^{\prime}%
}. \label{eq_Wkk}%
\end{align}
The driving potential in Eq. (\ref{eq_Wkk}) is given by
\begin{equation}
\hat{U}_{\boldsymbol{kk}^{\prime}}=-i\frac{\partial\delta\left(
\boldsymbol{k}-\boldsymbol{k}^{\prime}\right)  }{\partial\boldsymbol{k}}%
\cdot\hat{\boldsymbol{F}}_{\boldsymbol{k}^{\prime}} \label{eq_Ukkp}%
\end{equation}
with $\hat{\boldsymbol{F}}_{\boldsymbol{k}}=e^{-i\boldsymbol{k}\cdot
\boldsymbol{r}}\hat{\boldsymbol{F}}e^{i\boldsymbol{k}\cdot\boldsymbol{r}}$. The derivation of the current density operator is provided in the next section.

In momentum space, the retarded Green's function is defined as%
\begin{equation}
\langle\langle c_{\boldsymbol{k}}|c_{\boldsymbol{k}^{\prime}}^{\dag}%
\rangle\rangle_{t,t^{\prime}}^{r}=-\frac{i}{\hbar}\Theta(t-t^{\prime}%
)\langle\{c_{\boldsymbol{k}}(t),c_{\boldsymbol{k}^{\prime}}^{\dag}(t^{\prime
})\}\rangle,
\end{equation}
with its Fourier transform $G_{\boldsymbol{kk}^{\prime}}^{r}(\epsilon
)=\langle\langle c_{\boldsymbol{k}}|c_{\boldsymbol{k}^{\prime}}^{\dag}%
\rangle\rangle_{\epsilon}^{r}$ following the equation of
motion\cite{Deng_2016,PhysRevB.101.205137,PhysRevB.95.115102,PhysRevB.106.195134}%
\begin{equation}
\epsilon^{+}\langle\langle c_{\boldsymbol{k}}|c_{\boldsymbol{k}^{\prime}%
}^{\dag}\rangle\rangle_{\epsilon}^{r}=\langle\{c_{\boldsymbol{k}%
},c_{\boldsymbol{k}^{\prime}}^{\dag}\}\rangle+\langle\langle\left[
c_{\boldsymbol{k}},H\right]  |c_{\boldsymbol{k}^{\prime}}^{\dag}\rangle
\rangle_{\epsilon}^{r} \label{eq_GEOM}%
\end{equation}
where $\epsilon^{\pm}=\epsilon\pm i0^{+}$ and $\Theta(x)$ is the heaviside
function. For simplicity, we focus on the case of dilute doping and neglect
the impurity-impurity correlation. Combining Eqs.
(\ref{eq_Hkk}) and (\ref{eq_GEOM}), we can derive the Dyson equation%
\begin{equation}
G_{\boldsymbol{kk}^{\prime}}^{r}(\epsilon)=\mathcal{G}_{\boldsymbol{k}%
}(\epsilon)\delta_{\boldsymbol{kk}^{\prime}}+\mathcal{G}_{\boldsymbol{k}%
}(\epsilon)\sum_{\boldsymbol{l}}\hat{U}_{\boldsymbol{kl}}G_{\boldsymbol{lk}%
^{\prime}}^{r}(\epsilon), \label{eq_DyS}%
\end{equation}
where $\mathcal{G}_{\boldsymbol{k}}(\epsilon)=\left[  1-g_{\boldsymbol{k}%
}(\epsilon)\Sigma_{\boldsymbol{k}}(\epsilon)\right]  ^{-1}g_{\boldsymbol{k}%
}(\epsilon)$ and $g_{\boldsymbol{k}}(\epsilon)=\left(  \epsilon^{+}%
-\mathcal{H}_{\boldsymbol{k}}\right)  ^{-1}$ are the equilibrium retarded
Green's functions with and without impurity perturbation,
respectively. In Eq. (\ref{eq_DyS}), the Zeeman term $-\hat{\boldsymbol{M}%
}_{s}\cdot\boldsymbol{B}$ has been absorbed into the
self-energy\cite{PhysRevB.94.235116,PhysRevB.96.155141,PhysRevB.99.165146}%
\begin{equation}
\Sigma_{\boldsymbol{k}}(\epsilon)=[\mathbf{1}-\mathcal{\tilde{V}%
}_{\boldsymbol{k}}\frac{1}{N}\sum_{\boldsymbol{l}}g_{\boldsymbol{l}}%
(\epsilon)]^{-1}\mathcal{\tilde{V}}_{\boldsymbol{k}}, \label{eq_Tkk}%
\end{equation}
in which $\mathcal{\tilde{V}}_{\boldsymbol{k}}=\frac{1}{N}\sum_{\boldsymbol{k}%
^{\prime}}\langle\mathcal{V}_{\boldsymbol{kk}^{\prime}}\rangle_{\mathrm{dis}%
}-\hat{\boldsymbol{M}}_{s}\cdot\boldsymbol{B}$ and $\langle\mathcal{\cdots
}\rangle_{\mathrm{dis}}$ means the disordered average.

In momentum space, the ensemble average of an operator $\hat{Q}=\sum_{\boldsymbol{kk}^{\prime}%
}c_{\boldsymbol{k}}^{\dag}\hat{Q}_{\boldsymbol{kk}^{\prime}}c_{\boldsymbol{k}%
^{\prime}}$, can be computed using the retarded Green’s function
\begin{equation}
\langle\hat{Q}\rangle=-\frac{1}{\pi}\operatorname{Im}\int_{-\infty}^{\infty
}\sum_{\boldsymbol{kk}^{\prime}}\mathrm{Tr}\left[  \hat{Q}_{\boldsymbol{kk}%
^{\prime}}G_{\boldsymbol{k}^{\prime}\boldsymbol{k}}^{r}(\epsilon)\right]
f(\epsilon)d\epsilon\label{eq_Q}%
\end{equation}
where $f\left(  \epsilon\right)  =1/\left[  e^{(\epsilon-E_{F})/k_{B}%
T}+1\right]  $ is the Fermi-Dirac distribution function. In derivation of
Eq. (\ref{eq_Q}), the identity\cite{PhysRevB.99.085106}%
\begin{equation}
\langle c_{\boldsymbol{k}}^{\dag}c_{\boldsymbol{k}^{\prime}}\rangle
=-i\int_{-\infty}^{\infty}\frac{d\epsilon}{2\pi}G_{\boldsymbol{k}^{\prime
}\boldsymbol{k}}^{<}(\epsilon),
\end{equation}
has been employed, where $G_{\boldsymbol{k}^{\prime}\boldsymbol{k}}^{<}(\epsilon)=\left[
G_{\boldsymbol{k}^{\prime}\boldsymbol{k}}^{a}(\epsilon)-G_{\boldsymbol{k}%
^{\prime}\boldsymbol{k}}^{r}(\epsilon)\right]  f\left(  \epsilon\right)  $ is
the less Green's function and $G^{a}(\epsilon)=\left[  G^{r}(\epsilon)\right]
^{\dag}$ is the advanced Green's function. Substituting Eq. (\ref{eq_Ukkp})
into Eq. (\ref{eq_Q}) and integrating by parts, we can obtain%
\begin{equation}
\langle\hat{Q}\rangle=-\frac{1}{\pi}\operatorname{Im}\int_{-\infty}^{\infty
}\sum_{\boldsymbol{k}}\mathrm{Tr}\left[  \left(  \hat{Q}_{\boldsymbol{k}%
}+\Delta\hat{Q}_{\boldsymbol{k}}\right)  G_{\boldsymbol{k}}^{r}(\epsilon
)\right]  f(\epsilon)d\epsilon,
\end{equation}
where $\hat{Q}_{\boldsymbol{k}}=\hat{Q}_{\boldsymbol{kk}}$, $\Delta\hat
{Q}_{\boldsymbol{k}}=i\partial_{\boldsymbol{k}}\hat{Q}_{\boldsymbol{k}}%
\cdot\mathcal{G}_{\boldsymbol{k}}(\epsilon)\hat{\boldsymbol{F}}%
_{\boldsymbol{k}}$ and
\begin{equation}
G_{\boldsymbol{k}}^{r}(\epsilon)=\left[  1-i\partial_{\boldsymbol{k}%
}\mathcal{G}_{\boldsymbol{k}}(\epsilon)\cdot\hat{\boldsymbol{F}}%
_{\boldsymbol{k}}\right]  ^{-1}\mathcal{G}_{\boldsymbol{k}}(\epsilon)
\label{eq_GRk}%
\end{equation}
is the nonequilibrium retarded Green's function.

\begin{figure}[ptb]
\centering
\includegraphics[width=\linewidth]{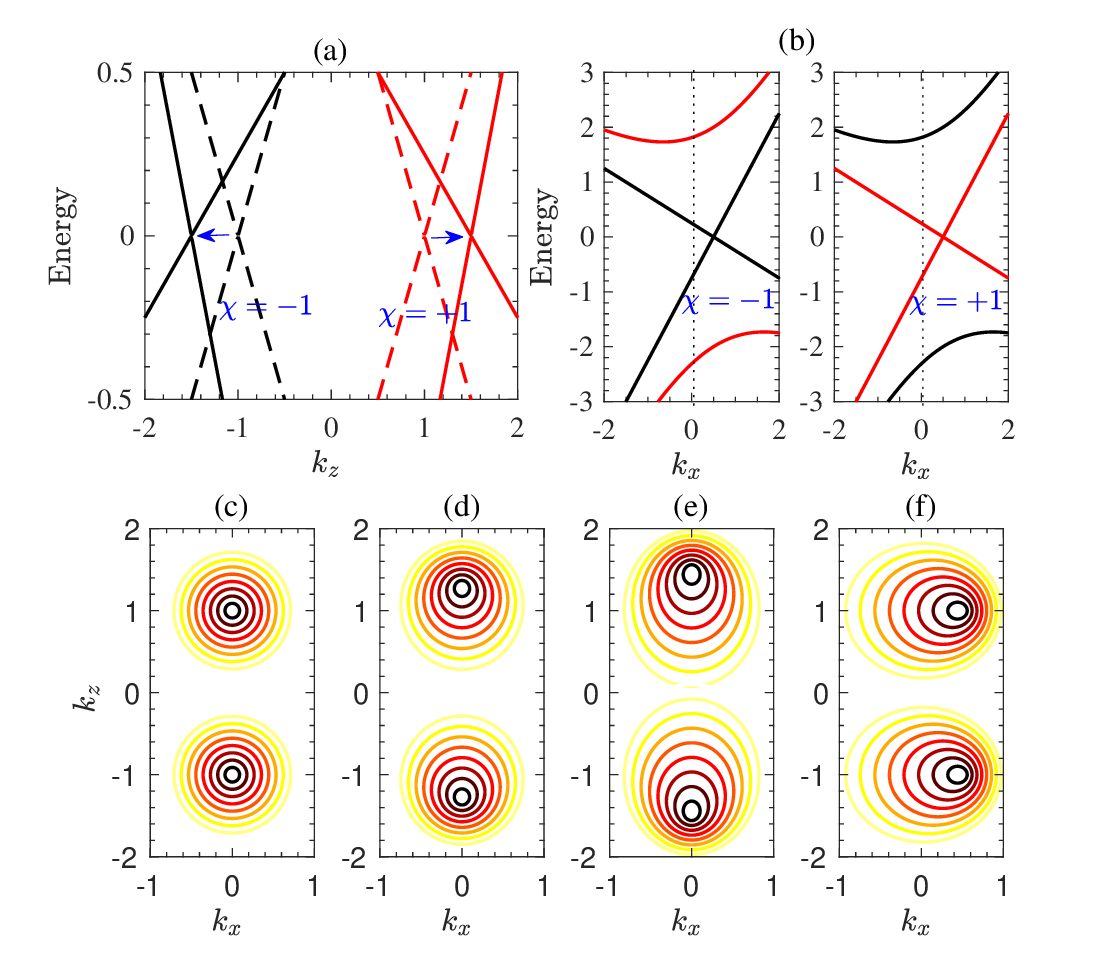}
\caption{(a)-(b) The Weyl cones tilted by the Zeeman field, with $B=0$ (dashed
lines) and $B=0.5$ (solid lines) for (a) oppositely tilted for $\boldsymbol{s}%
=\hat{e}_{z}$ and (b) identically tilted for $\boldsymbol{s}=\hat{e}_{x}$.
(c)-(f) Evolution of the energy contour with $\boldsymbol{s}=\hat{e}_{z}$ for
(c)-(e) $B=(0,0.3,0.5)$, and (f) $\boldsymbol{s}=\hat{e}_{x}$, $B=0.5$. The
rest parameters are set as $\hbar\upsilon_{\mathrm{F}}=1$, $\lambda=0.5$,
$k_{y}=0$ and $n_{\mathrm{i}}=0$.}%
\label{figEQ}%
\end{figure}

\section{Current density formula with driving fields}

\label{CDF}
In this section, we derive the current density within the framework of second quantization. To begin, the momentum and coordinate operators can be expressed as
\begin{align}
\hat{\boldsymbol{p}}  &  =\hbar\sum_{\boldsymbol{kk}^{\prime},\alpha
}\boldsymbol{k}\delta\left(  \boldsymbol{k}-\boldsymbol{k}^{\prime}\right)
c_{\boldsymbol{k}\alpha}^{\dag}c_{\boldsymbol{k}^{\prime}\alpha}%
,\label{eq_pk}\\
\hat{\boldsymbol{r}}  &  =i\sum_{\boldsymbol{kk}^{\prime},\alpha}%
\frac{\partial\delta\left(  \boldsymbol{k}-\boldsymbol{k}^{\prime}\right)
}{\partial\boldsymbol{k}}c_{\boldsymbol{k}\alpha}^{\dag}c_{\boldsymbol{k}%
^{\prime}\alpha}, \label{eq_rk}%
\end{align}
where $\alpha$ denotes the internal degrees of freedom of carriers and $\{c_{\boldsymbol{k}^{\prime}\alpha},c_{\boldsymbol{k}\beta}^{\dag}%
\}=\delta_{\boldsymbol{k}^{\prime}\boldsymbol{k}}\delta_{\alpha\beta}$. It can
be verified from Eqs. (\ref{eq_pk})-(\ref{eq_rk}) that $\left[  \hat{x}_{\mu
},\hat{p}_{\nu}\right]  =i\hbar\delta_{\mu\nu}\sum_{\boldsymbol{k},\alpha
}c_{\boldsymbol{k}\alpha}^{\dag}c_{\boldsymbol{k}\alpha}$ and%
\begin{equation}
\left[  \hat{x}_{\mu},\hat{x}_{\nu}\right]  =\sum_{\boldsymbol{k},\alpha
}\left(  \frac{\partial c_{\boldsymbol{k}\alpha}^{\dag}}{\partial k_{\mu}%
}\frac{\partial c_{\boldsymbol{k}\alpha}}{\partial k_{\nu}}-\frac{\partial
c_{\boldsymbol{k}\alpha}^{\dag}}{\partial k_{\nu}}\frac{\partial
c_{\boldsymbol{k}\alpha}}{\partial k_{\mu}}\right)  . \label{eq_xx}%
\end{equation}
For brevity, we define the Berry curvature operator $\hat
{\boldsymbol{\Omega}}_{\boldsymbol{k}}=-i\partial_{\boldsymbol{k}}^{\dag
}\times\partial_{\boldsymbol{k}}$, allowing the rewriting of Eq. (\ref{eq_xx}) as $\left[  \hat{x}_{\mu},\hat{x}_{\nu}\right]  =i\varepsilon
_{\mu\nu\rho}\sum_{\boldsymbol{k},\alpha}c_{\boldsymbol{k}\alpha}^{\dag}%
\hat{\Omega}_{\boldsymbol{k},\rho}c_{\boldsymbol{k}\alpha}$, where
$\varepsilon_{\mu\nu\rho}$ accounts for the Levi-Civita symbol. Henceforth, the Einstein summation convention is adopted for the directional indices.

Similarly, from Eqs. (\ref{eq_Hkk}) and (\ref{eq_rk}), we can determine
$\hat{\boldsymbol{J}}\equiv-iq\left[  \hat{\boldsymbol{r}},H\right]
/\hbar=\sum_{\boldsymbol{k}}c_{\boldsymbol{k}}^{\dag}\hat{\boldsymbol{J}%
}_{\boldsymbol{k}}c_{\boldsymbol{k}}$ with
\begin{equation}
\hat{\boldsymbol{J}}_{\boldsymbol{k}}=\frac{q}{\hbar}\left(  \frac
{\partial\mathcal{H}_{\boldsymbol{k}}}{\partial\boldsymbol{k}}-\hat
{\boldsymbol{F}}_{\boldsymbol{k}}\times\hat{\boldsymbol{\Omega}}%
_{\boldsymbol{k}}\right)  , \label{eq_JF}%
\end{equation}
where $\mathcal{\tilde{H}}_{\boldsymbol{k}}=\mathcal{H}_{\boldsymbol{k}%
}+\Sigma_{\boldsymbol{k}}(\epsilon)$ and the driving force
\begin{equation}
\hat{\boldsymbol{F}}_{\boldsymbol{k}}=-q\left(  \boldsymbol{\nabla}%
\Phi+\partial_{t}\boldsymbol{A}\right)  +\hat{\boldsymbol{J}}_{\boldsymbol{k}%
}\times\left(  \boldsymbol{\nabla}\times\boldsymbol{A}\right)  \label{eq_FR}%
\end{equation}
is derived from Eq. (\ref{eq_pp}). Combining Eqs. (\ref{eq_JF}) and
(\ref{eq_FR}), an iterative equation can be derived for the current density
operator%
\begin{equation}
\hat{\boldsymbol{J}}_{\boldsymbol{k}}=\frac{q\hat{D}_{\boldsymbol{k}}^{-1}%
}{\hbar}\left[  \frac{\partial\mathcal{H}_{\boldsymbol{k}}}{\partial
\boldsymbol{k}}-q\boldsymbol{E}\times\hat{\boldsymbol{\Omega}}_{\boldsymbol{k}%
}-\left(  \hat{\boldsymbol{J}}_{\boldsymbol{k}}\cdot\hat{\boldsymbol{\Omega}%
}_{\boldsymbol{k}}\right)  \boldsymbol{B}\right]  . \label{eq_JkD}%
\end{equation}
Then, substituting Eq. (\ref{eq_JkD}) into Eq. (\ref{eq_Q}) yields
\begin{equation}
\boldsymbol{J}=-\frac{1}{\pi}\operatorname{Im}\int_{-\infty}^{\infty}%
\sum_{\boldsymbol{k}}\mathrm{Tr}\left[  \left(  \hat{\boldsymbol{J}%
}_{\boldsymbol{k}}+\Delta\hat{\boldsymbol{J}}_{\boldsymbol{k}}\right)
G_{\boldsymbol{k}}^{r}(\epsilon)\right]  f(\epsilon)d\epsilon. \label{eq_JFR}%
\end{equation}
The phase-space factor $\hat{D}_{\boldsymbol{k}}=1-q\hat{\boldsymbol{\Omega}%
}_{\boldsymbol{k}}\cdot\boldsymbol{B}/\hbar$ in the current density operator
can be eliminated by replacing the summation over $\boldsymbol{k}$ with an
integral\cite{RevModPhys.82.1959}, i.e., $\sum_{\boldsymbol{k}}\rightarrow\frac{1}{\left(  2\pi\right)
^{3}}\int d^{3}\boldsymbol{k}\hat{D}_{\boldsymbol{k}}$.

Obviously, the expectation of $\dot{\boldsymbol{r}}=\hat{\boldsymbol{J}%
}_{\boldsymbol{k}}/q$ and $\dot{\boldsymbol{k}}=\hat{\boldsymbol{F}%
}_{\boldsymbol{k}}/\hbar$ can recover the semiclassical formula for the
electron wave packet dynamics\cite{RevModPhys.82.1959}. However, as different from the
semiclassical formula, Eq. (\ref{eq_JFR}) allows evaluation of the nonequilibrium electron transport without introducing phenomenological relaxation processes and incorporates the contribution of the Lorentz force in a nature way. To further demonstrate this, we apply our theory to driven disordered WSMs and investigate the second-order nonlinear magnetotransport.

\begin{figure}[ptb]
\centering
\includegraphics[width=\linewidth]{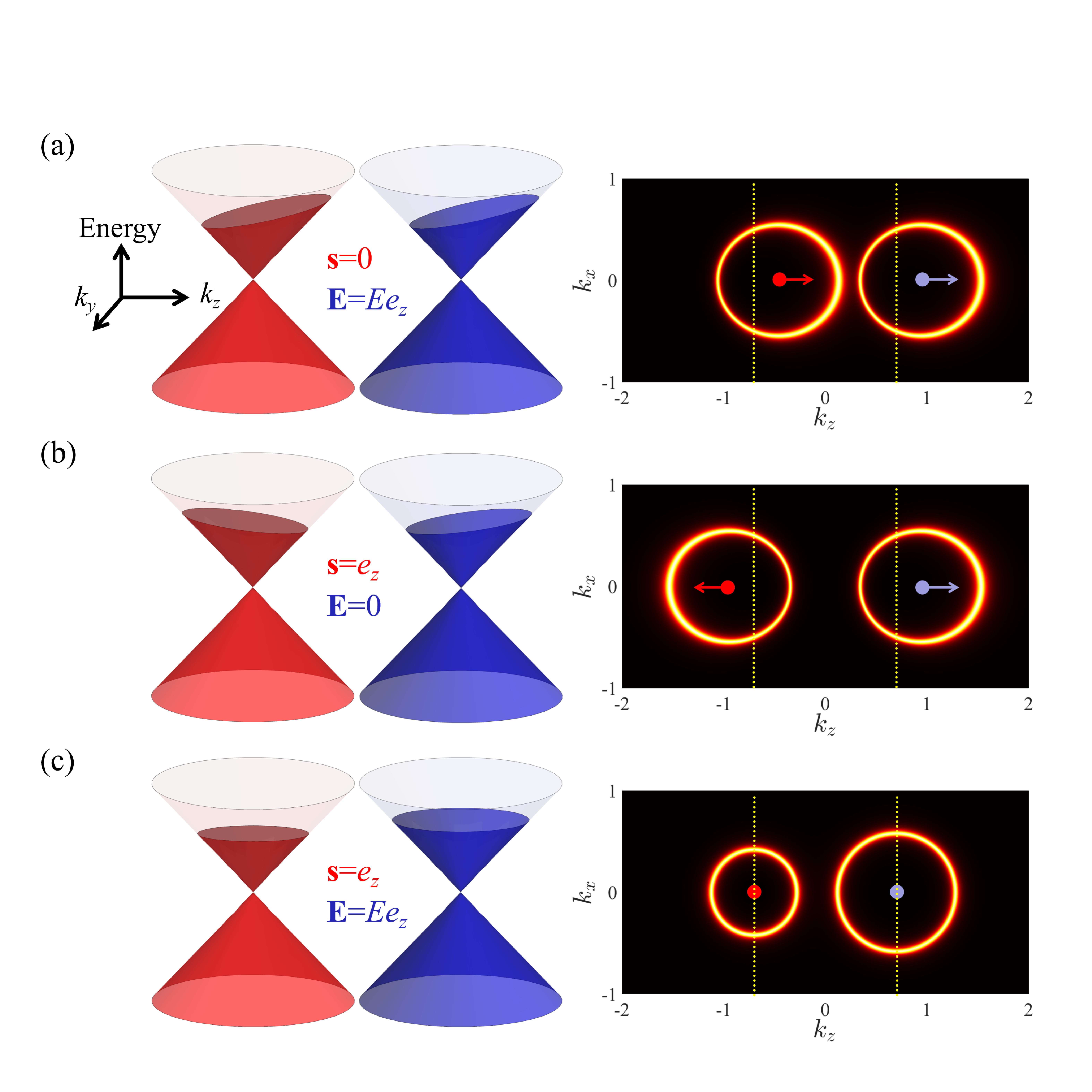}
\caption{The nonequilibrium population (left panel) and Fermi surface (right
panel) driven respectively by the (a) electric field, (b) Lorentz force and
(c) chiral anomaly.}%
\label{figDS}%
\end{figure}

\section{Nonequilibrium retarded Green's function for disordered WSMs}

\label{NRG}

We consider a WSM with a pair of Weyl nodes separated in the $z$ axis. Around
the Weyl nodes, the low-energy quasiparticle excitation can be described by
\begin{equation}
\mathcal{H}_{\boldsymbol{k}}^{\chi}=\lambda k^{2}+\hbar\upsilon_{\mathrm{F}%
}\boldsymbol{k}\cdot\boldsymbol{\sigma}^{\chi}, \label{eq_Hk}%
\end{equation}
where $\chi=\pm1$ denotes the chirality, $\boldsymbol{\sigma}^{\chi}=\left(
\sigma_{x},\sigma_{y},\chi\sigma_{z}\right)  $ is the spin Pauli matrix and
$\lambda$ characterizes the particle-hole asymmetry. Without external fields,
the impurity-free retarded Green's function takes the form%
\begin{align}
g_{\boldsymbol{k}}^{\chi}(\epsilon)  &  =\sum_{\eta}\frac{|\psi
_{\boldsymbol{k}\eta}^{\chi}\rangle\langle\psi_{\boldsymbol{k}\eta}^{\chi}%
|}{\epsilon^{+}-\varepsilon_{\boldsymbol{k}\eta}^{\chi}}\nonumber\\
&  =\frac{1}{2}\sum_{\eta}\frac{1}{\epsilon^{+}-\varepsilon_{\boldsymbol{k}%
\eta}^{\chi}}\left(  1+\eta\frac{\boldsymbol{k}}{k}\cdot\boldsymbol{\sigma
}^{\chi}\right)  , \label{eq_grk}%
\end{align}
where $\eta=\pm1$ is the helicity of the Weyl fermions and $|\psi
_{\boldsymbol{k}\eta}^{\chi}\rangle=\left(  \chi\cos\theta_{\boldsymbol{k}%
\eta}^{\chi},\sin\theta_{\boldsymbol{k}\eta}^{\chi}e^{i\phi_{\boldsymbol{k}}%
}\right)  ^{T}$ is the wavefuntion for the dispersion $\varepsilon
_{\boldsymbol{k}\eta}^{\chi}=\lambda k^{2}+\eta\hbar\upsilon_{\mathrm{F}}k$.
The related angles are defined as $\phi_{\boldsymbol{k}}=\tan^{-1}\left(
k_{y}/k_{x}\right)  $ and $\theta_{\boldsymbol{k}\eta}^{\chi}=\frac
{\theta_{\boldsymbol{k}}}{2}+\frac{1-\eta\chi}{4}\pi$ with $\theta
_{\boldsymbol{k}}=\cos^{-1}\left(  k_{z}/k\right)  $. Upon summing over $\boldsymbol{k}$, the last term in the
second line of Eq. (\ref{eq_grk}) vanishes, resulting in
\begin{equation}
g(\epsilon)=\frac{1}{N}\sum_{\boldsymbol{k}}g_{\boldsymbol{k}}^{\chi}%
(\epsilon)=\frac{\tilde{\epsilon}^{2}}{2\pi^{2}\hbar^{3}\upsilon_{\mathrm{F}%
}^{3}}\left(  \frac{1}{2}\ln\frac{\tilde{\epsilon}+\Lambda}{\tilde{\epsilon
}-\Lambda}-\frac{\Lambda}{\tilde{\epsilon}}\right)
\end{equation}
which is chirality-independent. Here, we define $\tilde{\epsilon}=\epsilon^{+}-\lambda\left(
\epsilon/\hbar\upsilon_{\mathrm{F}}\right)  ^{2}$ and introduce a higher-energy cutoff denoted by $\Lambda$
 for the dispersion. The unperturbed dispersion is illustrated by the dashed lines in Fig. \ref{figEQ}(a).

The disordered effect is modeled by a set of randomly-distributed
scattering potentials, consisting of a charge potential with strength
$V_{0}$ and a spin-dependent potential of strength $V_{M}$, i.e., $V_{n}%
=V_{0}\sigma_{0}+V_{M}\boldsymbol{s}_{n}\cdot\boldsymbol{\sigma}$, where
$\boldsymbol{s}_{n}$ is the direction vector of the local magnetic moment. The
Zeeman effect of the magnetic field  can be incorporated into the impurity potential by
replacing $V_{M}\boldsymbol{s}_{n}\rightarrow V_{M}\boldsymbol{s}_{n}%
-g\mu_{\mathrm{B}}\boldsymbol{B}$, where $g$ denotes the Lande factor, and $\mu_{\mathrm{B}}$ represents Bohr magneton. Utilizing Eq. (\ref{eq_Tkk}), we can derive the self-energy
\begin{equation}
\Sigma_{\boldsymbol{k}}(\epsilon)=\Gamma_{+}\left(  \epsilon\right)
-\Gamma_{-}\left(  \epsilon\right)  \boldsymbol{s}\cdot\boldsymbol{\sigma},
\label{eq_self}%
\end{equation}
where $\boldsymbol{s}$ is the mean direction of $g\mu_{\mathrm{B}%
}\boldsymbol{B}-V_{M}\boldsymbol{s}_{n}$ and%
\begin{equation}
\Gamma_{\pm}\left(  \epsilon\right)  =-\frac{1}{2}\left[  \frac{1}%
{g(\epsilon)-1/V_{+}}\pm\frac{1}{g(\epsilon)-1/V_{-}}\right]
\end{equation}
with $V_{\pm}=n_{\mathrm{i}}V_{0}\pm|g\mu_{\mathrm{B}}\boldsymbol{B}%
-n_{\mathrm{i}}V_{M}\langle\boldsymbol{s}_{n}\rangle_{\mathrm{dis}}|$ and
$n_{\mathrm{i}}$  representing the impurity concentration.

As a result, the impurity-perturbed equilibrium retarded Green's function
is given by
\begin{equation}
\mathcal{G}_{\boldsymbol{k}}(\epsilon)=\frac{1}{\epsilon^{+}-\Gamma_{+}\left(
\epsilon\right)  -\mathcal{H}_{\boldsymbol{k}}^{\chi}+\Gamma_{-}\left(
\epsilon\right)  \boldsymbol{s}\cdot\boldsymbol{\sigma}}. \label{eq_Gk0}%
\end{equation}
Upon application of driving fields, the nonequilibrium retarded Green's
function, as provided by Eq. (\ref{eq_GRk}), can be expressed as%
\begin{equation}
G_{\boldsymbol{k}}^{r}(\epsilon)=\frac{1}{\mathcal{G}_{\boldsymbol{k}}%
^{-1}(\epsilon)-\hat{\boldsymbol{M}}_{J}\cdot\boldsymbol{B}+e\boldsymbol{E}%
\cdot i\tilde{\partial}_{\boldsymbol{k}}}, \label{eq_GrN}%
\end{equation}
where $\tilde{\partial}_{\boldsymbol{k}}=\mathcal{G}_{\boldsymbol{k}}%
^{-1}(\epsilon)\partial_{\boldsymbol{k}}\mathcal{G}_{\boldsymbol{k}}%
(\epsilon)$ and $\hat{\boldsymbol{M}}_{J}=i\tilde{\partial}_{\boldsymbol{k}%
}\times\hat{\boldsymbol{J}}_{\boldsymbol{k}}^{\chi}$ is the magnetic moment
induced by the current density%
\begin{equation}
\hat{\boldsymbol{J}}_{\boldsymbol{k}}^{\chi}=\frac{-e\hat{D}_{\boldsymbol{k}%
}^{-1}}{\hbar}\left[  \frac{\partial\mathcal{H}_{\boldsymbol{k}}^{\chi}%
}{\partial\boldsymbol{k}}+e\boldsymbol{E}\times\hat{\boldsymbol{\Omega}%
}_{\boldsymbol{k}}-\left(  \hat{\boldsymbol{J}}_{\boldsymbol{k}}^{\chi}%
\cdot\hat{\boldsymbol{\Omega}}_{\boldsymbol{k}}\right)  \boldsymbol{B}\right]
. \label{eq_Jkc}%
\end{equation}

\section{Current-induced anomalous orbital magnetic moment}

\label{CAM}

From the nonequilibrium retarded Green's function, we can obtain the effective
Hamiltonian $\mathcal{H}_{\mathrm{eff}}\left(  \boldsymbol{k}\right)
=\epsilon_{R}^{+}-\left[  G_{\boldsymbol{k}}^{r}(\epsilon)\right]  ^{-1}$,
which can be further expressed as%
\begin{equation}
\mathcal{H}_{\mathrm{eff}}\left(  \boldsymbol{k}\right)  =\mathcal{H}%
_{\boldsymbol{k}}^{\chi}-\Gamma_{-}\left(  \epsilon\right)  \boldsymbol{s}%
\cdot\boldsymbol{\sigma}+\hat{\boldsymbol{M}}_{J}\cdot\boldsymbol{B}%
-e\boldsymbol{E}\cdot i\tilde{\partial}_{\boldsymbol{k}} \label{eq_Heff}.%
\end{equation}
Here, $\epsilon_{R}^{+}=\epsilon^{+}-\Gamma_{+}\left(  \epsilon\right)  $. As
$g(\epsilon)\rightarrow0$ and $\Gamma_{\pm}\left(  \epsilon\right)  =\left(
V_{+}\pm V_{-}\right)  /2$ for $\epsilon\rightarrow0$, the effective
magnetization $\Gamma_{-}\left(  \epsilon\right)  \boldsymbol{s}%
\cdot\boldsymbol{\sigma}$ from the self-energy will shift the Weyl nodes from
$\boldsymbol{k}=0$ to
\begin{equation}
\boldsymbol{k}_{\Gamma}^{\chi}=\frac{V_{+}-V_{-}}{2}\left(  s_{x},s_{y},\chi
s_{z}\right)  .
\end{equation}
Around the shifted Weyl nodes, the retarded Green's function becomes%
\begin{equation}
\mathcal{G}_{\boldsymbol{k}}(\epsilon)=\sum_{\eta}\frac{|\psi_{\boldsymbol{k}%
\eta}^{\chi}\rangle\langle\psi_{\boldsymbol{k}\eta}^{\chi}|}{\epsilon
^{+}-\Gamma_{+}\left(  \epsilon\right)  -\epsilon_{\boldsymbol{k}\eta}^{\chi}}
\label{eq_Grkc}%
\end{equation}
where $\epsilon_{\boldsymbol{k}\eta}^{\chi}=\eta\hbar\upsilon_{\mathrm{F}%
}k+2\lambda\boldsymbol{k}_{\Gamma}^{\chi}\cdot\boldsymbol{k}$ and the
wavevector is measured from $\boldsymbol{k}_{\Gamma}^{\chi}$. Consequently, the
Weyl cones with opposite chiralities will be tilted in opposite (identical)
directions when the Weyl nodes are shifted along the $z$ ($x$ or $y$)
direction, as displayed by Figs. \ref{figEQ}(a)-(b), along with that the
energy contours will evolve into a series of ellipses, see Figs.
\ref{figEQ}(c)-(f), which breaks the rotational symmetry.

Furthermore, as indicated by Eq. (\ref{eq_Jkc}), additional transport channels due to
the Berry curvature of the wavefunctions will be introduced by the external
fields. Excitation of band fermions to these chiral channels results in a decrease in the probability density of electrons on the energy band. Consequently, the normalization of the perturbed wavefunction will be less than or
equal to one, namely, $|\tilde{\psi}_{\boldsymbol{k}\eta}^{\chi}\rangle
=\sqrt{\mathcal{N}_{\chi}}|\psi_{\boldsymbol{k}\eta}^{\chi}\rangle$ with%
\begin{equation}
\mathcal{N}_{\chi}=\frac{\sum_{\boldsymbol{k}\eta}\delta\left(  \epsilon
-\epsilon_{\boldsymbol{k}\eta}^{\chi}\right)  }{\sum_{\boldsymbol{k}\eta
}\delta\left(  \epsilon-\epsilon_{\boldsymbol{k}\eta}^{\chi}\right)
+N_{\mathrm{ch}}}.
\end{equation}
The number of the chiral channels is determined
by\cite{PhysRevB.106.075139}%
\begin{equation}
N_{\mathrm{ch}}=\frac{1}{e\upsilon_{\mathrm{F}}}|\sum_{\boldsymbol{k}\eta
}\langle\psi_{\boldsymbol{k}\eta}^{\chi}|\hat{\boldsymbol{J}}_{\boldsymbol{k}%
}^{\chi}|\psi_{\boldsymbol{k}\eta}^{\chi}\rangle\delta\left(  \epsilon
-\epsilon_{\boldsymbol{k}\eta}^{\chi}\right)  |.
\end{equation}
It is evident that without the Berry curvature, $N_{\mathrm{ch}}=0$ and
$\mathcal{N}_{\chi}=1$, so that the zeroth order perturbation wavefunction for
a topologically trivial system remains unchanged, reverting to
ordinary perturbation theory. For the WSMs, the nonvanishing Berry
curvature%
\begin{equation}
\boldsymbol{\Omega}_{\boldsymbol{k}\eta}^{\chi}=\langle\psi_{\boldsymbol{k}%
\eta}^{\chi}|\hat{\boldsymbol{\Omega}}_{\boldsymbol{k}}|\psi_{\boldsymbol{k}%
\eta}^{\chi}\rangle=-\chi\eta\frac{\boldsymbol{k}}{2k^{3}}
\end{equation}
results in a finite $N_{\mathrm{ch}}=\frac{\omega_{c}^{2}}{4\pi^{2}\hbar\upsilon_{\mathrm{F}}^{3}%
}$ and $\mathcal{N}_{\chi}=2\epsilon^{2}/(2\epsilon^{2}%
+\hbar^{2}\omega_{c}^{2})$ deviating from unity. This implies that the fermions
can be pumped by a magnetic field from one valley to the other. Here, $\omega
_{c}=\upsilon_{\mathrm{F}}/\ell_{B}$ denotes the cyclotron frequency, and
$\ell_{B}=\sqrt{\hbar/eB}$ is the magnetic length.

\begin{figure}[ptb]
\centering
\includegraphics[width=\linewidth]{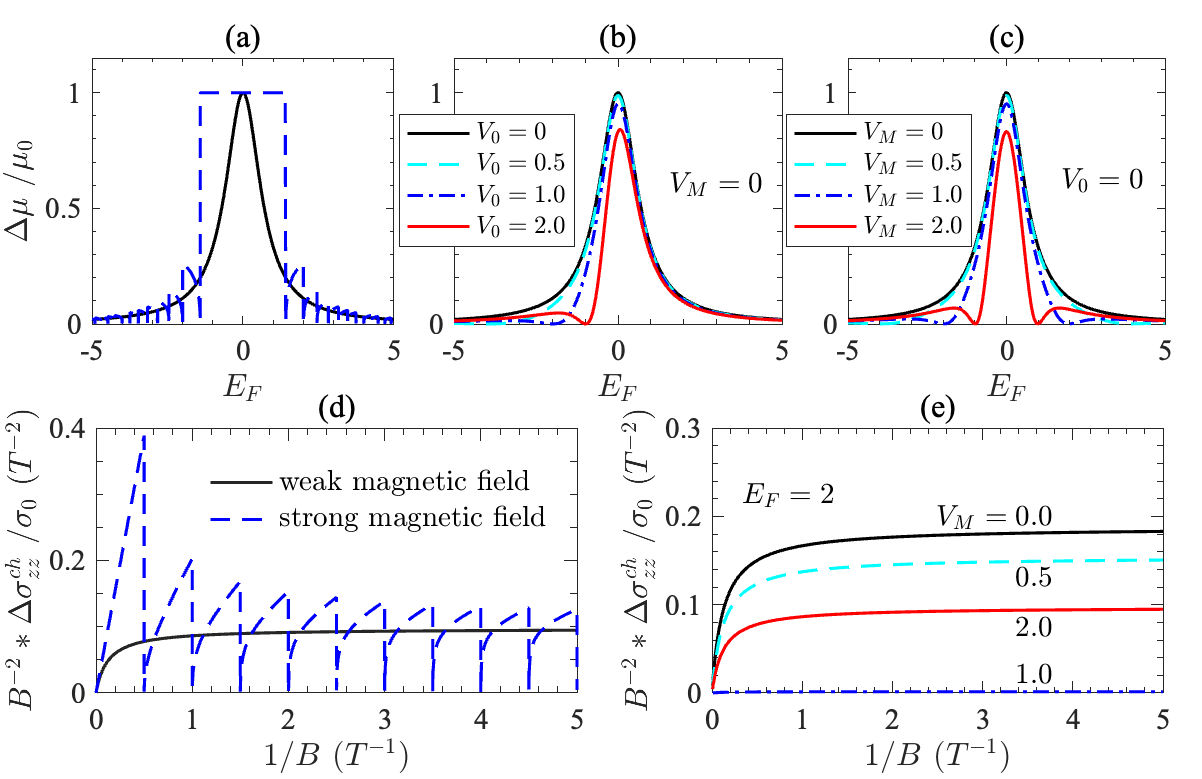}
\caption{(a)-(c) The chiral chemical potential $\Delta\mu$ vs the Fermi energy
$E_{F}$ with $\mu_{0}=eE\upsilon_{\mathrm{F}}\tau_{\mathrm{F}}|_{n_{\mathrm{i}%
}\rightarrow0}$ for (a) $n_{\mathrm{i}}=0$, (b) $n_{\mathrm{i}}=0.1$,
$V_{M}=0$, $V_{0}=(0,0.5,1.5,2)$ and (c) $n_{\mathrm{i}}=0.1$, $V_{0}=0$,
$V_{M}=(0,0.5,1.5,2)$. (c)-(e) The chiral-anomaly induced linear
magnetoconductivity $\Delta\sigma_{zz}^{\mathrm{ch}}(B)$ vs $1/B$ for varied
impurity scattering potential. The blue dashed lines in (a) and (d) are the
results taken into account the Landau level quantization, where a quantum
oscillation behavior is observable. Here, we set $\sigma_{0}=\Delta\sigma
_{zz}^{\mathrm{ch}}(B=1T)$, $\Lambda=100$ and the other parameters are the
same as Fig. \ref{figEQ}.}%
\label{figLC}%
\end{figure}

In the vicinity of the shifted Weyl nodes, the nonequilibrium spectrum is given by%
\begin{equation}
\tilde{\epsilon}_{\boldsymbol{k}\eta}^{\chi}=\epsilon_{\boldsymbol{k}\eta
}^{\chi}+\boldsymbol{M}_{J,\boldsymbol{k}\eta}^{\chi}\cdot\boldsymbol{B}%
-e\boldsymbol{E}\cdot\frac{\partial\epsilon_{\boldsymbol{k}\eta}^{\chi}}%
{\hbar\partial\boldsymbol{k}}\tau_{\boldsymbol{k}\eta}^{\chi},
\end{equation}
where $\tau_{\boldsymbol{k}\eta}^{\chi}=\hbar/\operatorname{Im}\left(
\epsilon_{R}^{+}\right)  |_{\epsilon\rightarrow\epsilon_{\boldsymbol{k}\eta
}^{\chi}}$ is the lifetime of the Weyl fermions and%
\begin{equation}
\boldsymbol{M}_{J,\boldsymbol{k}\eta}^{\chi}=\sum_{\eta^{\prime}}%
\frac{i\langle\tilde{\psi}_{\boldsymbol{k}\eta}^{\chi}|\partial
_{\boldsymbol{k}}\mathcal{H}_{\boldsymbol{k}}^{\chi}|\psi_{\boldsymbol{k}%
\eta^{\prime}}^{\chi}\rangle}{\epsilon_{\boldsymbol{k}\eta}^{\chi}%
-\epsilon_{\boldsymbol{k}\eta^{\prime}}^{\chi}+i\hbar/\tau_{\boldsymbol{k}%
\eta^{\prime}}^{\chi}}\times\langle\psi_{\boldsymbol{k}\eta^{\prime}}^{\chi
}|\hat{\boldsymbol{J}}_{\boldsymbol{k}}^{\chi}|\tilde{\psi}_{\boldsymbol{k}%
\eta}^{\chi}\rangle\label{eq_MJk}%
\end{equation}
is the expectation of $\hat{\boldsymbol{M}}_{J}$ in the renormalized state
$|\tilde{\psi}_{\boldsymbol{k}\eta}^{\chi}\rangle$. Substituting the first term
of Eq. (\ref{eq_Jkc}), i.e., $\hat{\boldsymbol{J}}_{\boldsymbol{k}}%
\rightarrow-e\partial_{\boldsymbol{k}}\mathcal{H}_{\boldsymbol{k}}/\hbar$,
into Eq. (\ref{eq_MJk}), we can obtain the Lorentz-force induced normal
orbital magnetic moment\cite{RevModPhys.82.1959}
\begin{equation}
\boldsymbol{M}_{J,\boldsymbol{k}\eta}^{(0)}=-\frac{ie}{\hbar}\mathcal{N}%
_{\chi}\langle\partial_{\boldsymbol{k}}\psi_{\boldsymbol{k}\eta}^{\chi}%
|\times\left(  \epsilon_{\boldsymbol{k}\eta}^{\chi}-\mathcal{H}%
_{\boldsymbol{k}}^{\chi}\right)  |\partial_{\boldsymbol{k}}\psi
_{\boldsymbol{k}\eta}^{\chi}\rangle.\label{eq_MJ0}%
\end{equation}
Analogously, replacing $\hat{\boldsymbol{J}}_{\boldsymbol{k}}%
\rightarrow-e^{2}\boldsymbol{E}\times\hat{\boldsymbol{\Omega}}_{\boldsymbol{k}%
}/\hbar$ in Eq. (\ref{eq_MJk}), we can obtain the current-induced anomalous
orbital magnetic moment%
\begin{equation}
\boldsymbol{M}_{J,\boldsymbol{k}\eta}^{(E)}=\frac{e^{2}}{\hbar}\mathcal{N}%
_{\chi}\left(  \boldsymbol{E}\times\boldsymbol{\Omega}_{\boldsymbol{k}\eta
}^{\chi}\right)  \times\frac{\partial\epsilon_{\boldsymbol{k}\eta}^{\chi}%
}{\hbar\partial\boldsymbol{k}}\tau_{\boldsymbol{k}\eta}^{\chi}.\label{eq_MJE}%
\end{equation}

As indicated by  Eq. (\ref{eq_MJ0}), the normal orbital
magnetic moment arises from the interband transition, i.e., $\eta^{\prime}%
\neq\eta$ in Eq. (\ref{eq_MJk}). On the other hand, the anomalous orbital magnetic moment  Eq. (\ref{eq_MJE})
originates from the intraband transition with $\eta^{\prime}=\eta$, which
depends on the Berry curvature and reflects the topology of the Fermi
surface. Without the electric field, $\boldsymbol{M}_{J,\boldsymbol{k}\eta
}^{\chi}=\boldsymbol{M}_{J,\boldsymbol{k}\eta}^{(0)}$ is the normal orbital
magnetic moment due to the cyclotron motion induced by the Lorentz force.
However, in systems with spin-orbit coupling, the spin and orbital magnetic
moments can couple to each other. Consequently, the anomalous term $\boldsymbol{M}_{J,\boldsymbol{k}\eta}^{(E)}$ arising from the Berry curvature is introduced. The interaction of the anomalous orbital magnetic moment with external fields can lead to novel physical phenomena, such as the chiral anomaly discussed below.
\begin{figure}[ptb]
\centering\includegraphics[width=\linewidth]{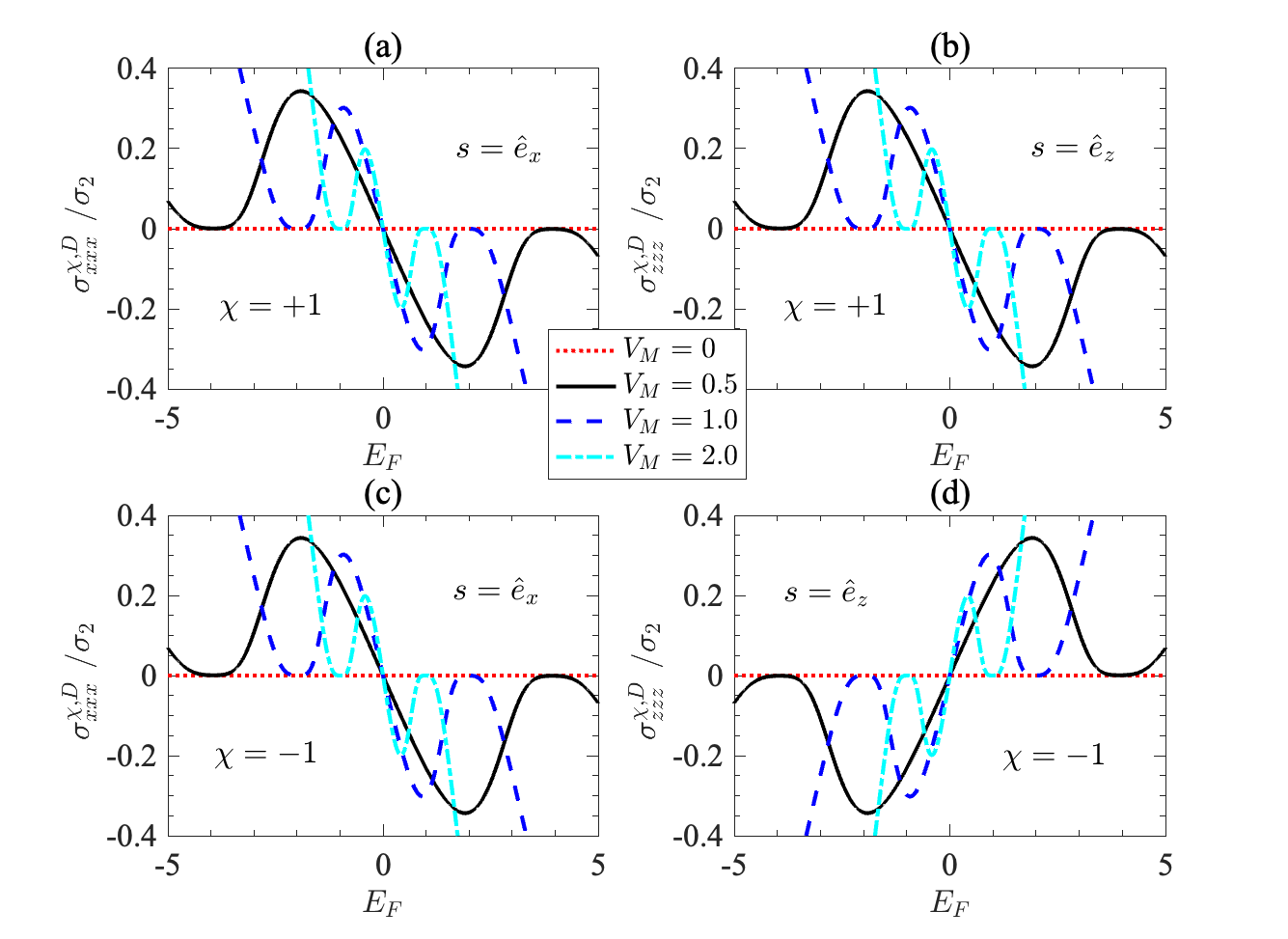}
\caption{The nonlinear longitudinal Drude conductivity $\sigma_{xxx}%
^{\chi,\mathrm{D}}$ and $\sigma_{zzz}^{\chi,\mathrm{D}}$ as functions of
$E_{F}$ with (a)-(b) $\chi=+$ and (c)-(d) $\chi=-$ for $V_{0}=0$ and
$V_{M}=(0,0.5,1,2)$, where the direction of the impurity moment is along the
$x$ ($z$) direction in the left (right) panel. The other parameters are the
same as Fig. \ref{figLC}.}%
\label{figND}%
\end{figure}

\section{Chiral anomaly originating from the anomalous orbital magnetic
moment}

\label{CFM} The chiral anomaly manifests itself through the change in the chemical potential for the Weyl valleys, quantified by the averaging of
\begin{equation}
\Delta\epsilon_{\boldsymbol{k}\eta}^{\chi}=\tilde{\epsilon}_{\boldsymbol{k}%
\eta}^{\chi}-\epsilon_{\boldsymbol{k}\eta}^{\chi}=\boldsymbol{M}%
_{J,\boldsymbol{k}\eta}^{\chi}\cdot\boldsymbol{B}-e\boldsymbol{E}%
\cdot\boldsymbol{\upsilon}_{\boldsymbol{k}\eta}^{\chi}\tau_{\boldsymbol{k}%
\eta}^{\chi}%
\end{equation}
over the states on the Fermi surface, i.e.,%
\begin{equation}
\Delta\mu_{\chi}=\frac{\sum_{\boldsymbol{k}\eta}\Delta\epsilon_{\boldsymbol{k}%
\eta}^{\chi}\delta\left(  E_{F}-\epsilon_{\boldsymbol{k}\eta}^{\chi}\right)
}{\sum_{\boldsymbol{k}\eta}\delta\left(  E_{F}-\epsilon_{\boldsymbol{k}\eta
}^{\chi}\right)  }.
\end{equation}
The orbital magnetic moments can be derived as%
\begin{align}
\boldsymbol{M}_{J,\boldsymbol{k}\eta}^{(0)}  &  =\chi\frac{2\ell_{B}%
^{2}e\upsilon_{\mathrm{F}}}{1+2\ell_{B}^{2}k^{2}}\boldsymbol{k}%
,\label{eq_MJk0}\\
\boldsymbol{M}_{J,\boldsymbol{k}\eta}^{(E)}  &  =\chi\eta\frac{e^{2}}{\hbar
}\frac{\ell_{B}^{2}\boldsymbol{\upsilon}_{\boldsymbol{k}\eta}^{\chi}%
\tau_{\boldsymbol{k}\eta}^{\chi}}{1+2\ell_{B}^{2}k^{2}}\times\left(
\boldsymbol{E}\times\frac{\boldsymbol{k}}{k}\right)  ,
\end{align}
where the band group velocity is given by
\begin{equation}
\boldsymbol{\upsilon}_{\boldsymbol{k}\eta}^{\chi}\equiv\frac{\partial
\epsilon_{\boldsymbol{k}\eta}^{\chi}}{\hbar\partial\boldsymbol{k}}=\left(
\frac{2\lambda\boldsymbol{k}_{\Gamma}^{\chi}}{\hbar\upsilon_{F}}+\eta
\frac{\boldsymbol{k}}{k}\right)  \upsilon_{\mathrm{F}}.
\end{equation}
While the Berry curvature is divergent at the Weyl nodes, $\boldsymbol{M}%
_{J,\boldsymbol{k}\eta}^{(0)}$ vanishes and $\boldsymbol{M}_{J,\boldsymbol{k}%
\eta}^{(E)}$ becomes a constant for $k=0$. Therefore, the nonequilibrium
spectrum is convergent in the whole Brillouin zone.

As specified by Eq. (\ref{eq_MJk0}), the Lorentz force can not induce the
chiral anomaly, because $\boldsymbol{M}_{J,\boldsymbol{k}\eta}^{(0)}%
=-\boldsymbol{M}_{J,-\boldsymbol{k}\eta}^{(0)}$ and $\boldsymbol{M}%
_{J,\boldsymbol{k}\eta}^{(0)}\cdot\boldsymbol{B}$ vanishes after averaging
over the Fermi surface. In contrast, $\boldsymbol{M}_{J,\boldsymbol{k}\eta
}^{(E)}$ contains a component even in $\boldsymbol{k}$ and when coupled with
$\boldsymbol{B}$, will alter the population in the Weyl valleys. In Fig.
\ref{figDS}, we plot the nonequilibrium population and Fermi surfaces by
inspecting the poles of the nonequilibrium retarded Green's function. As observed
in Figs. \ref{figDS}(a) and (b), the Fermi surfaces will be drifted by the
electric field (normal orbital magnetic moment induced by the Lorentz force) toward
the same (opposite) direction, while the chiral anomaly will change the size
of the Fermi surfaces, as depicted by Fig. \ref{figDS}(c). Concretely, the
nonequilibrium Fermi energy in the $\chi$ Weyl valley is $E_{F}^{\chi}%
=E_{F}+\chi\Delta\mu$, where the chiral chemical potential is given by%
\begin{equation}
\Delta\mu=\frac{e^{2}}{\hbar}\frac{\ell_{B}^{2}\upsilon_{\mathrm{F}}%
\tau_{\mathrm{F}}}{1+2\ell_{B}^{2}k_{F}^{2}}\boldsymbol{E}\cdot\boldsymbol{B}
\label{eq_Du}%
\end{equation}
with $\tau_{\mathrm{F}}=\hbar/\operatorname{Im}\left(  \epsilon_{R}%
^{+}\right)  |_{\epsilon\rightarrow E_{F}}$ and $k_{F}=E_{F}/\left(
\hbar\upsilon_{\mathrm{F}}\right)  $. Equation (\ref{eq_Du}) is consistent with previous derivations in both the
ultraquantum\cite{PhysRevLett.122.036601,PhysRevB.99.165146,QM_PhysRevB.107.L081107}
and semiclassical limits\cite{im_PhysRevB.103.045105,semi_PhysRevB.106.035423}.

As shown by Eq. (\ref{eq_Du}), $\Delta\mu\propto\boldsymbol{E}\cdot
\boldsymbol{B}/E_{F}^{2}$ for $\ell_{B}^{2}k_{F}^{2}\gg1$. In the
opposite regime, it converges to a constant, as indicated by the dark solid
line in Fig. \ref{figLC}(a). When taking the Landau level quantization into
account\cite{PhysRevB.106.075139}, i.e., $\ell_{B}^{2}k_{F}^{2}\rightarrow
\sum_{n=1}^{\ell_{B}^{2}k_{F}^{2}/2}\left[  1-2n/\left(  \ell_{B}^{2}k_{F}%
^{2}\right)  \right]  ^{-1/2}$, the chiral chemical potential will exhibit a
periodic-in-$1/B$ quantum oscillation, as illustrated by the blue dashed lines in
Fig. \ref{figLC}. Due to the impurity resonance states, the chiral chemical
potential experiences a dip at $g(\epsilon)V_{\pm}=1$. For nonmagnetic doping,
$V_{+}=V_{-}$ and only a resonant dip can be observed, as shown by Fig.
\ref{figLC}(b). However, in the case of magnetic doping, i.e., $V_{0}=0$ and
$V_{M}\neq0$, we have $V_{+}=-V_{-}$ and thus there appears a pair of resonant
dips distributed symmetrically about $E_{F}=0$, as illustrated by Fig.
\ref{figLC}(c). The symmetric structures of the resonant dips will be broken when both $V_{0}$ and $V_{M}$ are considered.
\begin{figure*}[ptb]
\centering\includegraphics[width=\linewidth]{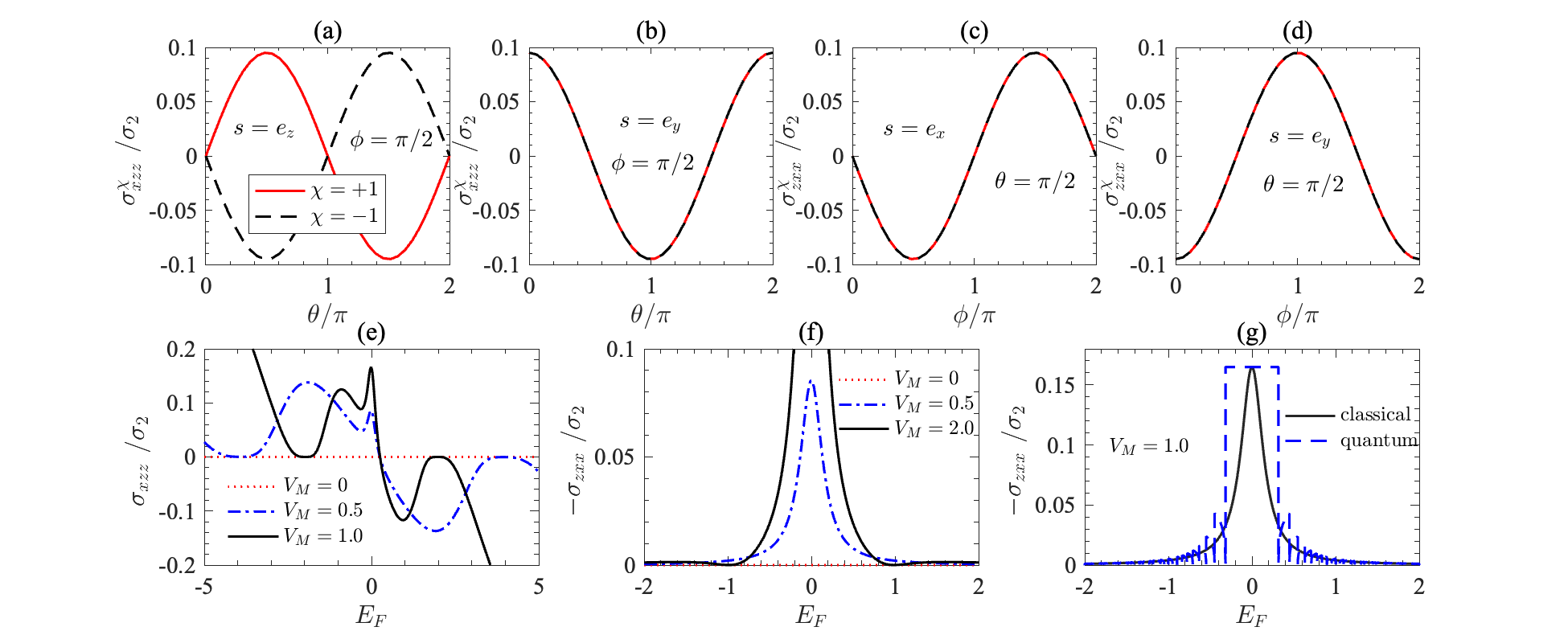}
\caption{The chirality-resolved nonlinear Hall conductivities (a)-(b)
$\sigma_{xzz}^{\chi}$ and (c)-(d) $\sigma_{zxx}^{\chi}$ as functions of the
polar $\theta$ and azimuth $\phi$ angles of the magnetic field with $E_{F}=0$
and $\boldsymbol{s}=\hat{e}_{z},\hat{e}_{y},\hat{e}_{x},\hat{e}_{y}$ in
sequence for (a)-(d). (e)-(f) The net nonlinear Hall conductivities
$\sigma_{xzz}$ and $\sigma_{zxx}$ vs the Fermi energy for $\boldsymbol{s}%
\parallel\hat{e}_{x}+\hat{e}_{y}$ and varied $V_{M}$. (g) Quantum
oscillation of the nonlinear Hall conductivity due to the chiral anomaly.
Here, we set $\lambda=1$, $B=0.05$ $V_{0}=0$ and other parameters the same as
Fig. \ref{figLC}.}%
\label{figNH}%
\end{figure*}

\section{Nonlinear magnetoconductivity in disordered WSMs}

\label{CBM} In this section, we estimate the conductivity based on the
nonequilibrium retarded Green's function. By substituting Eqs. (\ref{eq_GrN})
and (\ref{eq_Jkc}) into Eq. (\ref{eq_JFR}), the current density can be expressed as
\begin{equation}
\boldsymbol{J}^{\chi}=-\frac{1}{\pi}\operatorname{Im}\int_{-\infty}^{\infty
}\sum_{\boldsymbol{k}\eta}\frac{\langle\psi_{\boldsymbol{k}\eta}^{\chi}%
|\hat{\boldsymbol{J}}_{\boldsymbol{k}}^{\chi}+\Delta\hat{\boldsymbol{J}%
}_{\boldsymbol{k}}^{\chi}|\psi_{\boldsymbol{k}\eta}^{\chi}\rangle}%
{\epsilon^{+}-\Gamma_{+}\left(  \epsilon\right)  -\tilde{\epsilon
}_{\boldsymbol{k}\eta}^{\chi}}f(\epsilon)d\epsilon\label{eq_Jch}%
\end{equation}
with $\Delta\hat{\boldsymbol{J}}_{\boldsymbol{k}}^{\chi}=i\partial
_{\boldsymbol{k}}\hat{\boldsymbol{J}}_{\boldsymbol{k}}^{\chi}\cdot
\mathcal{G}_{\boldsymbol{k}}(\epsilon)\hat{\boldsymbol{F}}_{\boldsymbol{k}}$.
In the weak impurity scattering regime, i.e., $n_{\mathrm{i}}\ll1$, we can
approximate%
\begin{equation}
\operatorname{Im}\frac{1}{\epsilon^{+}-\Gamma_{+}\left(  \epsilon\right)
-\tilde{\epsilon}_{\boldsymbol{k}\eta}^{\chi}}=-\pi\delta\left(
\epsilon-\epsilon_{\boldsymbol{k}\eta}^{\chi}-\Delta\epsilon_{\boldsymbol{k}%
\eta}^{\chi}\right)  , \label{eq_Im}%
\end{equation}
such that Eq. (\ref{eq_Jch}) can be expanded in series of $\Delta
\epsilon_{\boldsymbol{k}\eta}^{\chi}$ as%
\begin{equation}
\boldsymbol{J}^{\chi}=\sum_{n=0}^{\infty}\frac{1}{n!}\int_{-\infty}^{\infty
}\left[  \boldsymbol{J}_{n}^{\chi}\left(  \epsilon\right)  +\Delta
\boldsymbol{J}_{n}^{\chi}\left(  \epsilon\right)  \right]  \frac{\partial
^{n}f(\epsilon)}{\partial\epsilon^{n}}d\epsilon\label{eq_JnE}%
\end{equation}
with%
\begin{equation}
\boldsymbol{J}_{n}^{\chi}\left(  \epsilon\right)  =\sum_{\eta}\int\frac
{d^{3}\boldsymbol{k}}{\left(  2\pi\right)  ^{3}}\delta\left(  \epsilon
-\epsilon_{\boldsymbol{k}\eta}^{\chi}\right)  \left(  \Delta\epsilon
_{\boldsymbol{k}\eta}^{\chi}\right)  ^{n}\boldsymbol{J}_{\boldsymbol{k}\eta
}^{\chi} \label{eq_Jcn}%
\end{equation}
and%
\begin{align}
\Delta\boldsymbol{J}_{n}^{\chi}\left(  \epsilon\right)   &  =\sum_{\eta}%
\int\frac{d^{3}\boldsymbol{k}}{\left(  2\pi\right)  ^{3}}\delta\left(
\epsilon-\epsilon_{\boldsymbol{k}\eta}^{\chi}\right)  \left(  \Delta
\epsilon_{\boldsymbol{k}\eta}^{\chi}\right)  ^{n}\nonumber\\
&  \times\frac{\partial\boldsymbol{J}_{\boldsymbol{k}\eta}^{\chi}}%
{\hbar\partial\boldsymbol{k}}\tau_{\boldsymbol{k}\eta}^{\chi}\cdot\left(
e\boldsymbol{E}-\boldsymbol{J}_{\boldsymbol{k}\eta}^{\chi}\times
\boldsymbol{B}\right)  .
\end{align}
For brevity, we adopted the abbreviation
\begin{equation}
\boldsymbol{J}_{\boldsymbol{k}\eta}^{\chi}=-e\left[  \boldsymbol{\upsilon
}_{\boldsymbol{k}\eta}^{\chi}+\frac{e}{\hbar}\boldsymbol{E}\times
\boldsymbol{\Omega}_{\boldsymbol{k}\eta}^{\chi}+\frac{e\boldsymbol{B}}{\hbar
}\left(  \boldsymbol{\upsilon}_{\boldsymbol{k}\eta}^{\chi}\cdot
\boldsymbol{\Omega}_{\boldsymbol{k}\eta}^{\chi}\right)  \right]  .
\label{eq_Jkec}%
\end{equation}
As we concentrate on the second-order nonlinear magnetotransport, the
current density can be truncated as%
\begin{equation}
J_{\mu}^{\chi}=\sigma_{\mu\nu}^{\chi}E_{\nu}+\sigma_{\mu\nu\rho}^{\chi}E_{\nu
}E_{\rho}+\mathcal{O}(\boldsymbol{E}^{3}),
\end{equation}
where $\sigma_{\mu\nu}^{\chi}$ and $\sigma_{\mu\nu\gamma}^{\chi}$ are the
linear and second-order nonlinear conductivities. Apparently, the current
density resulting from the second-order conductivity is invariant when
reversing the direction of the electric field and thus leads to the
nonreciprocal transport.

\subsection{Linear conductivity}

Before delving into the analysis of nonlinear conductivity, we first present the results for linear conductivity. Following a detailed algebraic procedure, we obtain a unified analytical expression for the linear conductivity tensor, which can be partitioned into three distinct components:
\begin{equation}
\sigma_{\mu\nu}^{\chi}=\sigma_{D}\delta_{\mu\nu}+\Delta\sigma_{\mu\nu
}^{\mathrm{ch}}(B)+\Delta\sigma_{\mu\nu}^{\mathrm{tilt}}(B),
\end{equation}
where $\sigma_{D}=n_{e}e^{2}\ell_{\mathrm{F}}/\left(  \hbar k_{F}\right)  $ is
the Drude conductivity and%
\begin{align}
\Delta\sigma_{\mu\nu}^{\mathrm{ch}}(B)  &  =\frac{e^{2}}{h}\frac{2\ell_{B}%
^{2}\ell_{\mathrm{F}}}{1+2\ell_{B}^{2}k_{F}^{2}}\frac{e^{2}B_{\mu}B_{\nu}%
}{2\pi\hbar^{2}}\left(  1-|\boldsymbol{\alpha}|^{2}\right)  ^{2}%
\label{eq_ch}\\
\Delta\sigma_{\mu\nu}^{\mathrm{tilt}}(B)  &  =\frac{\Delta n_{e}e^{2}%
\ell_{\mathrm{F}}}{\hbar k_{F}}-\frac{2e^{2}\ell_{\mathrm{F}}k_{F}^{2}%
}{1+2\ell_{B}^{2}k_{F}^{2}}\frac{\alpha_{\mu}B_{\nu}+\alpha_{\nu}B_{\mu}}{\pi
hB} \label{eq_tilt}%
\end{align}
are the linear magnetoconductivities from the chiral anomaly and tilting of
the Weyl cones, respectively. The related factors are defined as
$\boldsymbol{\alpha}=2\lambda\boldsymbol{k}_{\Gamma}^{\chi}/\left(
\hbar\upsilon_{\mathrm{F}}\right)  $ and $\Delta n_{e}=\tilde{n}_{e}%
-n_{e}\delta_{\mu\nu}$, in which $n_{e}=\left(  1/3\pi^{2}\right)  k_{F}^{3}$
is the carrier density, $\ell_{\mathrm{F}}=\upsilon_{\mathrm{F}}%
\tau_{\mathrm{F}}$ denotes the mean free-path of the Weyl fermions and%
\begin{equation}
\tilde{n}_{e}=\frac{3n_{e}}{2\left(  1-|\boldsymbol{\alpha}|^{2}\right)  ^{2}%
}\left[  \left(  \frac{2-3\zeta}{|\boldsymbol{\alpha}|^{2}}-2\right)
\alpha_{\mu}\alpha_{\nu}+\zeta\delta_{\mu\nu}\right]  \label{eq_ne}%
\end{equation}
with $\zeta=\frac{1-|\boldsymbol{\alpha}|^{2}}{|\boldsymbol{\alpha}|^{2}%
}\left(  1-\frac{1-|\boldsymbol{\alpha}|^{2}}{2|\boldsymbol{\alpha}|}\ln
\frac{1+|\boldsymbol{\alpha}|}{1-|\boldsymbol{\alpha}|}\right)  $.

The linear magnetoconductivities are displayed in Figs. \ref{figLC}(d) and
(e). From Eq. (\ref{eq_ch}), we see that the linear magnetoconductivity
induced by the chiral anomaly scales with $B^{2}/k_{F}^{2}$ when $\ell
_{B}^{2}k_{F}^{2}\gg1$. Conversely, for $\ell_{B}^{2}k_{F}^{2}\ll1$, $\Delta
\sigma_{\mu\nu}^{\mathrm{ch}}(B)$ becomes independent of $k_{F}$ and
linear in $B$. These outcomes align with predictions from both semiclassical and quantum Boltzmann theories\cite{PhysRevLett.122.036601,PhysRevB.99.165146,QM_PhysRevB.107.L081107,im_PhysRevB.103.045105,semi_PhysRevB.106.035423}. Moreover,  the Zeeman field induced tilting of the Weyl cones can
contribute a magnetoconductivity $\propto B^{2}k_{F}^{2}$, as demonstrated by
the first term of Eq. (\ref{eq_tilt}). The combined effect of the tilting and the
Berry curvature results in the second term of Eq. (\ref{eq_tilt}), which has
been reported to be $k_{F}$%
-independent\cite{PhysRevB.99.115121,PhysRevB.104.L121117}. Here, we find this
conclusion holds true only for large Fermi levels. In the limit $k_{F}\rightarrow0$,
this term vanishes due to the diminishing density of states. All the linear
magnetoconductivities exhibit a linear dependence on the lifetime of the Weyl fermions, leading to significant suppression around the impurity resonance states.

\subsection{Impurity-scattering modulated bilinear magnetoconductivity}

For the sake of discussion, we refer to $e\boldsymbol{E}\times
\boldsymbol{\Omega}_{\boldsymbol{k}\eta}^{\chi}/\hbar$ and $e\boldsymbol{B}%
\left(  \boldsymbol{\upsilon}_{\boldsymbol{k}\eta}^{\chi}\cdot
\boldsymbol{\Omega}_{\boldsymbol{k}\eta}^{\chi}\right)  /\hbar$ in Eq.
(\ref{eq_Jkec}) as the anomalous velocity and pumping velocity, respectively.
Here, the nonvanishing second-order conductivity can can arise from several mechanisms, which to linear order in $B$ can be expressed as%
\begin{equation}
\sigma_{\mu\nu\rho}^{\chi}=\sigma_{\mu\nu\rho}^{\chi,\mathrm{A}}+\sigma
_{\mu\nu\rho}^{\chi,\mathrm{D}}+\Delta\sigma_{\mu\nu\rho}^{\chi,\mathrm{L}%
}+\Delta\sigma_{\mu\nu\rho}^{\chi,\mathrm{P}}+\Delta\sigma_{\mu\nu\rho}%
^{\chi,\mathrm{M}}.
\end{equation}
To elaborate, the term $\sigma_{\mu\nu\rho}^{\chi,\mathrm{A}}\sim\left(  \boldsymbol{E}%
\cdot\boldsymbol{\upsilon}_{\boldsymbol{k}\eta}^{\chi}\right)  \left(
\boldsymbol{E}\times\boldsymbol{\Omega}_{\boldsymbol{k}\eta}^{\chi}\right)
_{\mu}$ represents the nonlinear anomalous Hall conductivity originating from $\boldsymbol{J}%
_{1}^{\chi}\left(  \epsilon\right)  $ in Eq. (\ref{eq_JnE}). The nonlinear
Drude conductivity $\sigma_{\mu\nu\rho}^{\chi,\mathrm{D}}$ comprises two
contributions:  $\left(  \boldsymbol{E}\cdot\boldsymbol{\upsilon
}_{\boldsymbol{k}\eta}^{\chi}\right)  ^{2}\upsilon_{\mu,\boldsymbol{k}\eta
}^{\chi}$ arising from $\boldsymbol{J}_{2}^{\chi}\left(  \epsilon\right)  $ and
$\left(  \boldsymbol{E}\cdot\boldsymbol{\upsilon}_{\boldsymbol{k}\eta}^{\chi
}\right)  \boldsymbol{E}\cdot\partial_{\boldsymbol{k}}\upsilon_{\mu
,\boldsymbol{k}\eta}^{\chi}$ arising from $\Delta\boldsymbol{J}_{1}^{\chi}\left(
\epsilon\right)  $. Both  the nonlinear anomalous Hall and Drude conductivities are not explicit functions of magnetic field.
The remaining terms represent the nonlinear magnetoconductivities explicitly
involving both electric and magnetic fields. Specifically,
\begin{align}
\Delta\sigma_{\mu\nu\rho}^{\chi,\mathrm{P}}  &  \sim\left(  \boldsymbol{E}%
\cdot\boldsymbol{\upsilon}_{\boldsymbol{k}\eta}^{\chi}\right)  \left[
\boldsymbol{E}\cdot\partial_{\boldsymbol{k}}\left(  \boldsymbol{\upsilon
}_{\boldsymbol{k}\eta}^{\chi}\cdot\boldsymbol{\Omega}_{\boldsymbol{k}\eta
}^{\chi}\right)  B_{\mu}\right] \\
\Delta\sigma_{\mu\nu\rho}^{\chi,\mathrm{L}}  &  \sim\left(  \boldsymbol{E}%
\cdot\boldsymbol{\upsilon}_{\boldsymbol{k}\eta}^{\chi}\right)  \left(
\boldsymbol{\upsilon}_{\boldsymbol{k}\eta}^{\chi}\times\boldsymbol{B}\right)
\cdot\partial_{\boldsymbol{k}}\left(  \boldsymbol{E}\times\boldsymbol{\Omega
}_{\boldsymbol{k}\eta}^{\chi}\right)  _{\mu}%
\end{align}
arise from the pumping velocity and the combined effect of the anomalous
velocity and Lorentz force in $\Delta\boldsymbol{J}_{1}^{\chi}\left(
\epsilon\right)  $, respectively. Additionally,  $\Delta\sigma_{\mu\nu\rho}%
^{\chi,\mathrm{M}}=\Delta\sigma_{\mu\nu\rho}^{\chi,\mathrm{NM}}+\Delta
\sigma_{\mu\nu\rho}^{\chi,\mathrm{AM}}$, where $\Delta\sigma_{\mu\nu\rho}%
^{\chi,\mathrm{NM}}\sim\left(  \boldsymbol{M}_{J,\boldsymbol{k}\eta}%
^{(0)}\cdot\boldsymbol{B}\right)  \left(  \boldsymbol{E}\cdot
\boldsymbol{\upsilon}_{\boldsymbol{k}\eta}^{\chi}\right)  \left(
\boldsymbol{E}\times\boldsymbol{\Omega}_{\boldsymbol{k}\eta}^{\chi}\right)
_{\mu}$ is attributable to the joint effect of the normal orbital magnetic moment
and anomalous velocity in $\boldsymbol{J}_{2}^{\chi}\left(  \epsilon\right)  $, and $\Delta\sigma_{\mu\nu\rho}^{\chi,\mathrm{AM}}$ arises from the anomalous
orbital magnetic moment consisting of a term $\sim\left(  \boldsymbol{M}%
_{J,\boldsymbol{k}\eta}^{(E)}\cdot\boldsymbol{B}\right)  \left(
\boldsymbol{E}\times\boldsymbol{\Omega}_{\boldsymbol{k}\eta}^{\chi}\right)
_{\mu}$ from $\boldsymbol{J}_{1}^{\chi}$ and a term $\sim$ $\left(
\boldsymbol{M}_{J,\boldsymbol{k}\eta}^{(E)}\cdot\boldsymbol{B}\right)
\boldsymbol{E}\cdot\partial_{\boldsymbol{k}}\upsilon_{\mu,\boldsymbol{k}\eta
}^{\chi}$ from $\Delta\boldsymbol{J}_{1}^{\chi}\left(  \epsilon\right)  $. Consequently, the chiral anomaly contributes to the nonlinear transport by interacting either with the anomalous velocity or the derivative of the band
group velocity.

In particular, upon integrating the above expressions over $\boldsymbol{k}$ and
$\epsilon$, the nonlinear anomalous Hall conductivity can be expressed as
$\sigma_{\mu\nu\rho}^{\chi,\mathrm{A}}=\chi\sigma_{2}\varepsilon_{\mu\nu\rho
}/2$ and the nonlinear Drude conductivity, to leading order in
$\boldsymbol{\alpha}$, takes the form%
\begin{equation}
\sigma_{\mu\nu\rho}^{\chi,\mathrm{D}}=-\sigma_{2}\frac{E_{F}\tau_{\mathrm{F}}%
}{\hbar}\left(  \alpha_{\mu}\delta_{\nu\rho}+\alpha_{\nu}\delta_{\rho\mu
}+3\alpha_{\rho}\delta_{\mu\nu}\right)  ,
\end{equation}
where $\sigma_{2}=\frac{e^{3}\tau_{\mathrm{F}}}{6\pi^{2}\hbar^{2}%
}|_{n_{\mathrm{i}}\rightarrow0}$ is set as the unit of the second-order
conductivity. Similarly, the nonlinear magnetoconductivities originating from
the pumping velocity and Lorentz force can be summarized as%
\begin{align}
\Delta\sigma_{\mu\nu\rho}^{\chi,\mathrm{P}}  &  =-\chi\sigma_{2}^{\prime}%
\frac{E_{F}\tau_{\mathrm{F}}}{\hbar}\frac{2B_{\mu}\delta_{\nu\rho}}%
{B}\label{eq_sigmaP}\\
\Delta\sigma_{\mu\nu\rho}^{\chi,\mathrm{L}}  &  =-\chi\sigma_{2}^{\prime}%
\frac{E_{F}\tau_{\mathrm{F}}}{\hbar}\frac{B_{\mu}\delta_{\nu\rho}-B_{\nu
}\delta_{\mu\rho}}{B} \label{eq_sigmaL}%
\end{align}
with $\sigma_{2}^{\prime}=\sigma_{2}/\left(  1+2\ell_{B}^{2}k_{F}^{2}\right)
$. The nonlinear magnetoconductivities contributed by the normal and anomalous
orbital magnetic moments can be expressed as%
\begin{align}
\Delta\sigma_{\mu\nu\rho}^{\chi,\mathrm{NM}}  &  =-\sigma_{2}^{\prime}%
\frac{\alpha_{\rho}B_{\gamma}}{B}\varepsilon_{\mu\nu\gamma},\label{eq_sigmaNM}%
\\
\Delta\sigma_{\mu\nu\rho}^{\chi,\mathrm{AM}}  &  =-\sigma_{2}^{\prime}%
\frac{\alpha_{\gamma}B_{\rho}-\alpha_{\rho}B_{\gamma}}{2B}\varepsilon_{\mu
\nu\gamma}\nonumber\\
&  +\chi\sigma_{2}^{\prime}\frac{2E_{F}\tau_{\mathrm{F}}}{\hbar}\frac{B_{\mu
}\delta_{\nu\rho}+B_{\nu}\delta_{\rho\mu}+B_{\rho}\delta_{\mu\nu}}{5B}.
\label{eq_sigmaAM}%
\end{align}
Combining Eqs. (\ref{eq_sigmaP})-(\ref{eq_sigmaAM}), the total
nonlinear magnetoconductivity is given by%
\begin{align}
\Delta\sigma_{\mu\nu\rho}^{\chi}  &  =-\sigma_{2}^{\prime}\frac{\alpha_{\rho
}B_{\gamma}+\alpha_{\gamma}B_{\rho}}{2B}\varepsilon_{\mu\nu\gamma}\nonumber\\
&  -\chi\sigma_{2}^{\prime}\frac{E_{F}\tau_{\mathrm{F}}}{\hbar}\frac{13B_{\mu
}\delta_{\nu\rho}-7B_{\nu}\delta_{\mu\rho}-2B_{\rho}\delta_{\mu\nu}}{5B}.
\label{eq_dsigma}%
\end{align}

In the context of a single Weyl cone, we observe three primary characteristics in the second-order nonlinear conductivity. ($\mathrm{i}$) The nonlinear anomalous Hall conductivity and the nonlinear Hall conductivity induced by the orbital magnetic moment exhibit a linear scaling with the Fermi lifetime, while the nonlinear Drude conductivity and other magnetoconductivities are quadratic in $\tau_{\mathrm{F}}$. ($\mathrm{ii}$) Non-vanishing quadratic-in-$\tau_{\mathrm{F}}$ nonlinear magnetoconductivity necessitates at least two identical subscripts, while linear-in-$\tau_{\mathrm{F}}$ magnetoconductivities require at least two different subscripts. ($\mathrm{iii}$\textrm{) }The chiral anomaly due
to the Berry-curvature induced anomalous magnetic moment contributes both
the linear- and quadratic-in-$\tau_{\mathrm{F}}$ nonlinear
magnetoconductivity, as demonstrated by Eq. (\ref{eq_sigmaAM}). For the
former (latter),  the contribution is a constant (vanishing) as $E_{F}\rightarrow0$, evolving into a $B/E_{F}^{2}$ ($B/E_{F}$) dependency for $\ell_{B}^{2}k_{F}^{2}\gg1$. This is in contrast to the $B^{2}/E_{F}^{2}$ dependence observed in the linear case. These features align with the results obtained from the semiclassical Boltzmann theory\cite{im_PhysRevB.103.045105,semi_PhysRevB.106.035423}.

All components of the nonlinear conductivities that are independent of $\boldsymbol{\alpha}$ vanish upon summing over $\chi$ due to the inversion symmetry. Consequently, the dominant contributions to the net nonlinear conductivity arise from
\begin{equation}
\sigma_{\mu\nu\rho}(\boldsymbol{B})=\sum_{\chi}\left(  \sigma_{\mu\nu\rho
}^{\chi,\mathrm{D}}-\sigma_{2}^{\prime}\frac{\alpha_{\rho}B_{\gamma}%
+\alpha_{\gamma}B_{\rho}}{2B}\varepsilon_{\mu\nu\gamma}\right)  .
\label{eq_Dsigma}%
\end{equation}
In Fig. \ref{figND}, we display the numerical results for the nonlinear
longitudinal conductivities. For magnetic doping along the $z$ direction,
i.e., $\alpha_{z}\neq0$, the Weyl cones will tilt oppositely, see Fig.
\ref{figEQ} (a), which can not break the inversion symmetry. As a consequence,
the nonlinear conductivities from opposite Weyl valleys are of opposite signs
and effectively cancel each other out, as evident in Figs. \ref{figND}(b) and (d). Conversely, when the impurity moments align along the $x$ or $y$ direction, leading to identical tilting of the Weyl cones, the nonlinear conductivities from opposite Weyl valleys are additive, as illustrated in Figs. \ref{figND}(a) and (c). This results in an observable nonlinear longitudinal conductivity. 
Similar observations apply to the nonlinear Hall conductivities, as demonstrated in Figs. \ref{figNH}(a)-(d).For $E_{F}=0$, $\sigma_{\mu\nu\rho}^{\chi
,\mathrm{D}}$ vanishes and only the nonlinear Hall conductivity remains, with its
magnitude proportional directly to $V_{M}$. The nonlinear conductivities
are quite sensitive to impurity scattering and exhibit resonant dips
around the impurity resonance states.

According to Eq. (\ref{eq_Dsigma}), we find that $\sigma_{zzz}=0$ and
\begin{equation}
\sigma_{xxx}=-\frac{5e^{3}\tau_{\mathrm{F}}^{2}}{6\pi^{2}\hbar^{3}}E_{F}%
\alpha_{x},
\end{equation}
and the net nonlinear Hall conductivities are given by%
\begin{align}
\sigma_{zxx}  &  =-\frac{e^{3}\tau_{\mathrm{F}}}{6\pi^{2}\hbar^{2}}\frac
{1}{1+2\ell_{B}^{2}k_{F}^{2}}\frac{\alpha_{x}B_{y}+\alpha_{y}B_{x}}%
{B},\label{eq_sxzz}\\
\sigma_{xzz}  &  =-\frac{e^{3}\tau_{\mathrm{F}}}{6\pi^{2}\hbar^{2}}\left(
\frac{2\alpha_{x}}{\hbar}E_{F}\tau_{\mathrm{F}}-\frac{1}{1+2\ell_{B}^{2}%
k_{F}^{2}}\frac{\alpha_{y}B_{z}}{B}\right)  . \label{eq_szxx}%
\end{align}
While the linear conductivity adheres to the Onsager's relation
$\sigma_{\mu\nu}(\boldsymbol{B})=\sigma_{\nu\mu}(-\boldsymbol{B})$, the
nonlinear conductivity, as indicated by Eqs. (\ref{eq_sxzz}) and (\ref{eq_szxx}%
), violates the Onsager's relation, i.e., $\sigma_{\mu\nu\nu
}(\boldsymbol{B})\neq\sigma_{\nu\mu\mu}(-\boldsymbol{B})$,leading to
nonreciprocal transport behavior. With an increase in $k_{F}$, $\sigma
_{\mu\nu\rho}^{\chi,\mathrm{D}}$ becomes dominant and surpasses the last term of Eq. (\ref{eq_Dsigma}) at a critical point. 
Consequently, in addition to the resonant dips, an extra valley emerges in the nonlinear Hall conductivity $\sigma_{xzz}$, as depicted in Fig. \ref{figNH}(e), while the symmetric structure of $\sigma_{zxx}$ remains unchanged, as shown in Fig. \ref{figNH}(f).  If we consider Landau level quantization, the nonlinear Hall conductivity will exhibit synchronous quantum oscillations with the chiral chemical potential, as demonstrated in Fig. \ref{figNH}(g).

\section{Summary}
In summary, we have developed a nonlinear transport theory based on nonequilibrium retarded Green's functions, allowing us to investigate the nonlinear magnetotransport in disordered WSMs. Our study revealed that, for a single Weyl valley, the bilinear magnetoconductivity can be attributed to various mechanisms, including the normal orbital magnetic moment induced by the Lorentz force, the anomalous orbital magnetic moment induced by Berry curvature, and the impurity-induced tilting of the Weyl cones. Due to impurity scattering, the chiral anomaly experiences significant suppression around impurity-induced resonance states, leading to resonant dips in the magnetoconductivity when the Fermi energy approaches these states. Our results are consistent with the predictions from both the semiclassical and quantum Boltzmann theories.

\section{Acknowledgement}

This work was supported by the National NSF of China under Grants No.
12274146, No. 11874016, No. 12174121 and No. 12104167, the Guangdong Basic and
Applied Basic Research Foundation under Grant No. 2023B1515020050, and the
Guangdong Provincial Key Laboratory under Grant No. 2020B1212060066.

\bibliographystyle{apsrev4-1}
\bibliography{bibHMW}

\end{document}